\documentclass[twocolumn]{aastex6}

\shorttitle{Peer-review under review - Part I}
\shortauthors{Ferdinando Patat}

\begin{document}


\title{Peer-review under review \\ a statistical study on proposal ranking at ESO. Part I: the pre-meeting phase}


\author{Ferdinando Patat\altaffilmark{1}}
\affil{European Southern Observatory\\
K. Schwarzschildstr. 2 \\
D-85748 Garching b. M\"unchen, Germany}


\altaffiltext{1}{fpatat@eso.org}

\begin{abstract}
Peer review is the most common mechanism in place  for assessing requests for resources in a large variety of scientific disciplines. One of the strongest criticisms to this paradigm is the limited reproducibility of the process, especially at largely oversubscribed facilities. In this and in a subsequent paper we address this specific aspect in a quantitative way, through a statistical study on proposal ranking at the European Southern Observatory. For this purpose we analysed a sample of about 15000 proposals, submitted by more than 3000 Principal Investigators over 8 years. The proposals were reviewed by more than 500 referees, who assigned over 140000 grades in about 200 panel sessions. After providing a detailed analysis of the statistical properties of the sample, the paper presents an heuristic model based on these findings, which is then used to provide quantitative estimates of the reproducibility of the pre-meeting process.  On average, about one third of the proposals ranked in the top quartile by one referee are ranked in the same quartile by any other referee of the panel. A similar value is observed for the bottom quartile. In the central quartiles, the agreement fractions are very marginally above the value expected for a fully aleatory process (25\%). The agreement fraction between two panels composed by 6 referees is 55$\pm$5\% (50\% confidence level) for the top and bottom quartiles. The corresponding fraction for the central quartiles is 33$\pm$5\%. The model predictions are confirmed by the results obtained from boot-strapping the data for sub-panels composed by 3 referees, and fully consistent with the NIPS experiment. The post-meeting phase will be presented and discussed in a forthcoming paper.
\end{abstract}


\submitted{Draft version \today}



\section{Introduction} \label{sec:intro}

Peer review is the most popular mechanism in place  for assessing requests for resources in a large variety of scientific disciplines. In its most essential formulation, the peer review concept is based on the idea that referees with a comparable level of competence in the given field can provide an expert evaluation, which not only guarantees quality, but also maintains credibility.
This mechanism is deployed for publishing papers, distributing funds and allocating time on scientific facilities, be these particle accelerators, super-computers or telescopes, both ground-based or space-born.

Although the peer review paradigm has been object of strong criticism (see for instance \citet{smith} and \citet{gillies}), it is still seen as a democratic, self-regulating process which, notwithstanding its limitations, has no valid substitute.

One of the most common critiques by applicants at largely oversubscribed facilities is related to the stochastic component allegedly inherent to the process, which is claimed to be very significant. A quantitative analysis of this specific aspect is the main topic of this work. Although it concentrates on the specific case of telescope time requests at the European Southern Observatory (ESO), the present study was conducted aiming at deriving results of general applicability.

\begin{table*}
\begin{center}
\tabcolsep 5mm
\caption{\label{tab:grades} Grade scale}
\begin{tabular}{lll}
\hline
1.0 & outstanding & breakthrough science \\
1.5 & excellent  & definitely above average\\
2.0 & very good & no significant weaknesses\\
2.5 & good & minor deficiencies do not detract from strong scientific case\\
3.0 & fair & good scientific case, but with definite weaknesses\\
3.5 & rather weak & limited science return prospects\\
4.0 & weak & little scientific value and/or questionable scientific strategy\\
4.5 & very weak & deficiencies outweigh strengths\\
5.0 & rejected & \\
\hline
\end{tabular}
\end{center}
\end{table*}

ESO is an intergovernmental organisation and runs one of the largest ground-based astronomical facilities world-wide. Every semester, about 900 proposals including more than 3000 distinct scientists are submitted to ESO, requesting time on a large suite of telescopes: the four 8.2m units of the Very Large Telescope (VLT), VISTA (4.2m), 3.6m, NTT (3.5m), VST (2.6m) and the Atacama Pathfinder Experiment (APEX), placed on three different sites in Chile. 
The scale and the homogeneous way in which the telescope time applications have been reviewed at ESO during the last fifteen years provide a valuable database, which can be used to characterise the proposal review in a statistically robust way.

Given the complexity of the problem, we decided to split the study into two parts, separating the pre- and the post-meeting phases. This paper reports the results for the first phase of the process, while the second phase  will be presented in a forthcoming publication (hereafter Paper II), which will include the overall conclusions of the study.

The article is structured as follows. After giving a brief overview of the review process at ESO in Section~\ref{sec:process}, we present the data set in Section~\ref{sec:data}, and the general statistical properties of the pre-meeting grades in Section~\ref{sec:pregrades}. The distribution of average run grades is discussed in Section~\ref{sec:rungrades}, while Section~\ref{sec:corr} deals with the study of referee and panel correlations and agreement fractions. In Section~\ref{sec:model} we describe an heuristic approach for modelling the pre-meeting review process based on the statistical properties derived in this study. The model is validated in Section~\ref{sec:validation}, and used to quantify the reproducibility of the pre-meeting process in Section~\ref{sec:simul}.
We finally discuss the results in Section~\ref{sec:disc}, and summarize our conclusions in Section~\ref{sec:conc}.

\section{Proposal review at ESO}
\label{sec:process}

The proposal review process at ESO is described in detail in \cite{PH13}. Here only a summary of the parts relevant to the discussion will be given.

\subsection{Programme types}

In the current implementation, the following programme types are offered to the users:
Normal, Large, Target of Opportunity, Guaranteed Time Observations, Monitoring, Calibration and Director Discretionary Time (DDT). DDTs are reviewed by an internal committee, and a final decision is taken by the Director General.

Most of the proposals ($>$80\%) are of the Normal type, while ESO receives on average $\sim$20 Large Programme proposals each semester.

\subsection{The panels}

The proposals are distributed to 13 panels, which cover 4 science categories:  A: Cosmology (3 panels); B: Galaxies and galactic nuclei (2 panels); C: ISM, star formation and planetary systems (4 panels), and  D: Stellar evolution (4 panels). The different number of panels within each category reflects the different number of proposals these categories receive each period. Each panel has 6 members, including one panel chair and one panel co-chair.
The OPC proper is composed by the 13 panel chairs, 3 panel co-chairs (1 in A, 2 in B), and the OPC chair, who is not a panel member. This sums up to a total of 17 OPC members and 72 panel members, for a total of 79 scientists. OPC and panel members are selected on the basis of their scientific profile. The OPC members serve for 2 years (4 ESO periods), while panel members serve for 1 year (2 ESO periods). A fraction of the panel members are invited to serve a third semester, to ensure some level of continuity in the review process.

\subsection{The proposal review process}
\label{sec:review}

The proposal review process at ESO is split into two phases: an at-home, asynchronous review and a face-to-face meeting. The two steps are described in the following sub-sections.

\subsubsection{Pre-meeting phase}
\label{sec:pre}

The OPC and panels meet twice a year, about 50 days after the proposal submission deadlines. After a first check of the proper scientific category (followed by possible category re-assignments), the proposals are distributed to the panels members taking into account institutional conflicts. The referees are then given about one week to report scientific conflicts, and to request changes in the scientific category proposal assignments. Once this is completed, the refereeing process is started.
All panel members read all proposals assigned to their panel, and grade each run of these proposals\footnote{In the current implementation, users can apply for time using different runs within the same proposal. These may be used for requesting time at different instruments/telescopes, with different setups, etc.}. A scale between 1 (best) and 5 is used: the general meaning of the various bins is presented in Table~\ref{tab:grades}. For each run, each reviewer submits a single numerical grade.

The reviewers are given about four weeks to complete the review process, during which they enter the grades using a web interface. The grading is secret, meaning that single referees do not have access to the grades given by the other panel members.
Once this is done, the grades of all referees are normalised, so that the distribution of the grades of each of them has the same mean and standard deviation (see also Section~\ref{sec:norm}). The average normalised grade of the runs is used to compile ranked lists per telescope.

\subsubsection{Triage}

Following the increasing workload on the panels, ESO deployed a triage procedure to limit the number of proposals to be discussed at the meeting. The cumulative requested time per telescope is computed down each list and a triage line is drawn when this cumulative time exceeds 70\% of the total requested time on the considered telescope. As a rule, proposals below the triage line are not considered further. However, proposals for which the spread of the individual referee grades exceeds a certain threshold are brought back above the line. In addition, triaged proposals can be re-considered upon request of any panel member at the meeting.

\subsubsection{The meeting}
\label{sec:postmeeting}

The meeting is divided into a number of sessions, which involve the OPC proper and the panels. In the panel sessions, for each proposal the primary referee (randomly selected) gives a short presentation of the proposal under discussion, and presents her/his evaluation. All other (non-conflicted) panel members present their assessment. Finally, after a general discussion, secret vote takes place. For this, each panel member fills a voting slip with her/his acronym, the proposal identifier, and a grade (in the same scale used in the pre-meeting phase). The panel assistant collects the voting slips and enters the grades in the ESO database. The average and standard deviation of the individual grades are computed and assigned to the proposal. Although the grades given by the individual referees are stored in the database, the link between the grade and the referee is not stored in electronic format (only on the paper voting slips). 
The goal of the panel meetings is to discuss and grade all non-triaged proposals of Normal, Target of Opportunity and Guaranteed Time programmes. Large programmes go through a different process (for instance, they are not graded by the panels) and, therefore, they were not included in this analysis.

\begin{figure}
\centerline{ 
\includegraphics[width=9cm]{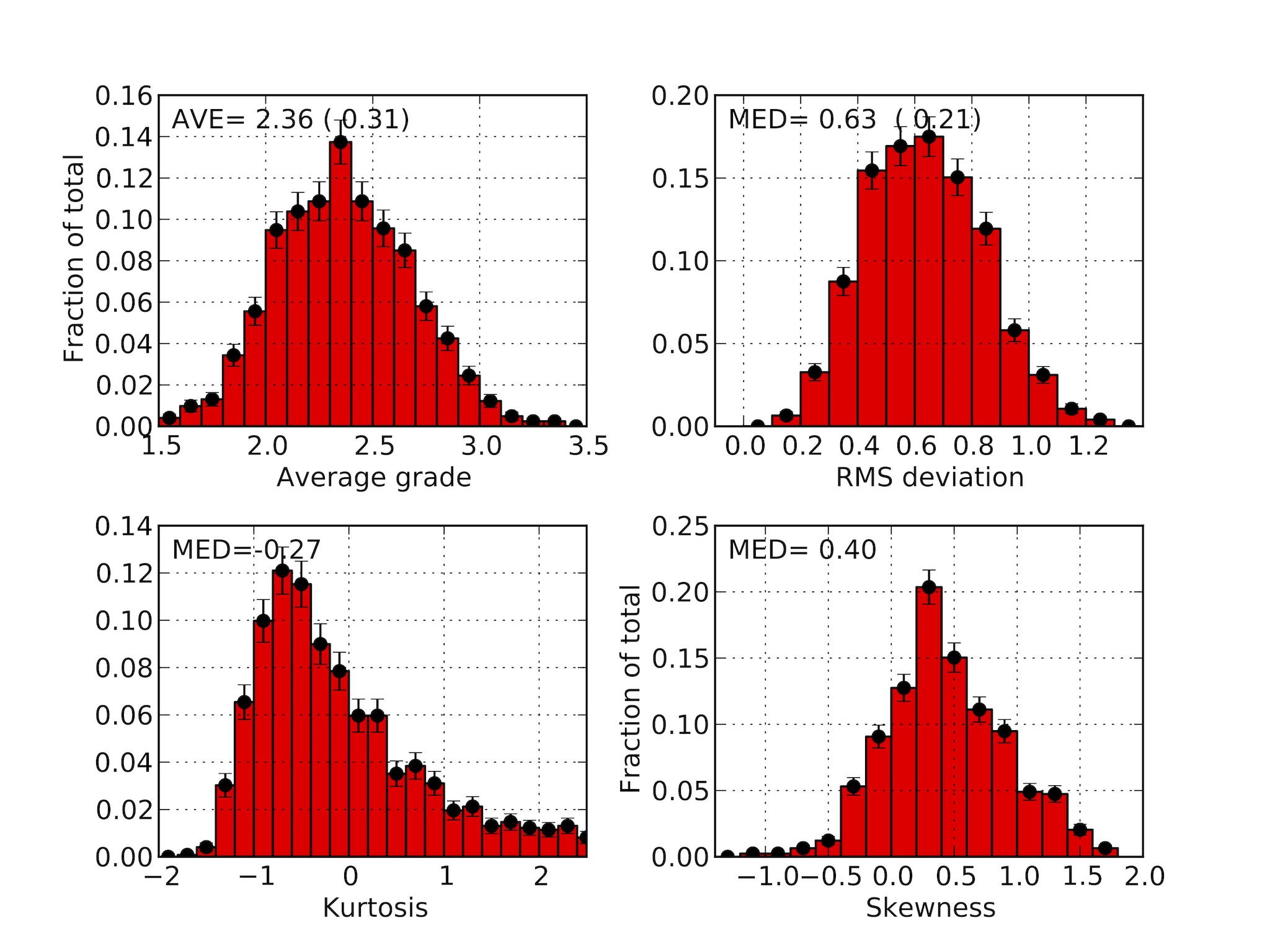}
}
\caption{\label{fig:summary}Distribution of statistical estimators for all the pre-meeting referee sessions. The error-bars indicate the Poissonian uncertainties.}
\end{figure}

In the current implementation, all runs with final average grades larger than 3.0 are not considered for scheduling and are, at all effects, rejected. The referees are aware of this formal cutoff when they review the proposals, so that a grading above 3.0 implies a deliberate decision of rejecting the given run.

\subsection{Referee calibration}
\label{sec:norm}

As we will see (Section~\ref{sec:data}), different referees can have very different grade
distributions. In particular, they may use different minimum-maximum
grade ranges, or have particularly generous or stingy attitudes. For
this reason, before merging the grades given to a run by different
referees, one may want to bring the various referees to the same
scale. This operation is indicated as referee calibration.

There are many possible ways to approach this problem. Here we limit
the discussion to the one that is currently implemented at ESO. 

Let us indicate with $\bar{g_j}$ and $\sigma_j$ the average grade and
the standard deviation of the $j$-th referee, and with $\langle g\rangle$ and
$\langle\sigma\rangle$ the mean average and mean standard deviation of all
referees. The original grade $g_j$ is calibrated via the following
transformation:

\begin{displaymath}
g^\prime _j= \langle g\rangle + \frac{\langle\sigma\rangle}{\sigma_j} \; \left (g_j - \bar{g_j} \right )
\end{displaymath}

This is a simple shift-and-stretch transformation: once applied, all
referee session distributions have the same average $\langle g\rangle$ and
standard deviation $\langle\sigma\rangle$. In the current ESO implementation the
calibration is computed using only grades within a selected range
(typically between 1.0 and 2.99) and only for selected programme
types. Since this is done on a semester-by-semester basis, only the
grades of the given period are used.

Grade calibration is always matter of discussion among the
referees, as there is no general agreement on the reasons/ways for/of applying
it. For instance, one can envisage more robust estimators for the
central value and the dispersion, and alternative transformations. In addition, the 
above prescription works under the assumption that the grade distributions are 
Gaussian-like. Although this is typically the case, deviations  from Normal behaviours 
are observed (see Section~\ref{sec:pregrades}).

Since adopting the specific calibration method deployed at ESO would produce a loss of generality,
all the analysis presented in this paper was conducted using raw (i.e. uncalibrated) grades. 
The effects of calibration are discussed in Appendix~\ref{sec:calib}.

\section{The data set}
\label{sec:data}

The data on which this study is based were extracted from the ESO database OPC70. The sample excludes the following programme types: DDT, Calibration, Large (which are not graded by the panels) and public surveys (which are reviewed by a special board).

\subsection{Period range}

The OPC70 database contains the pre- and post-meeting proposal grading starting with Period 79. However, the data are properly and consistently stored only starting with P82. For this reason, the analysis presented here covers ESO periods 82 to 97 (October 2018 - September 2016). This interval can be considered as representative of regular operations, with the VLT/VLTI in full activity and with practically all foci occupied by an instrument.
During P82 and 83 the OPC included 12 panels. An extra panel was added to category A in P84. From P84 to 87 the OPC had a stable composition of 13 panels, with six members each.

\begin{figure}
\centerline{ 
\includegraphics[width=9cm]{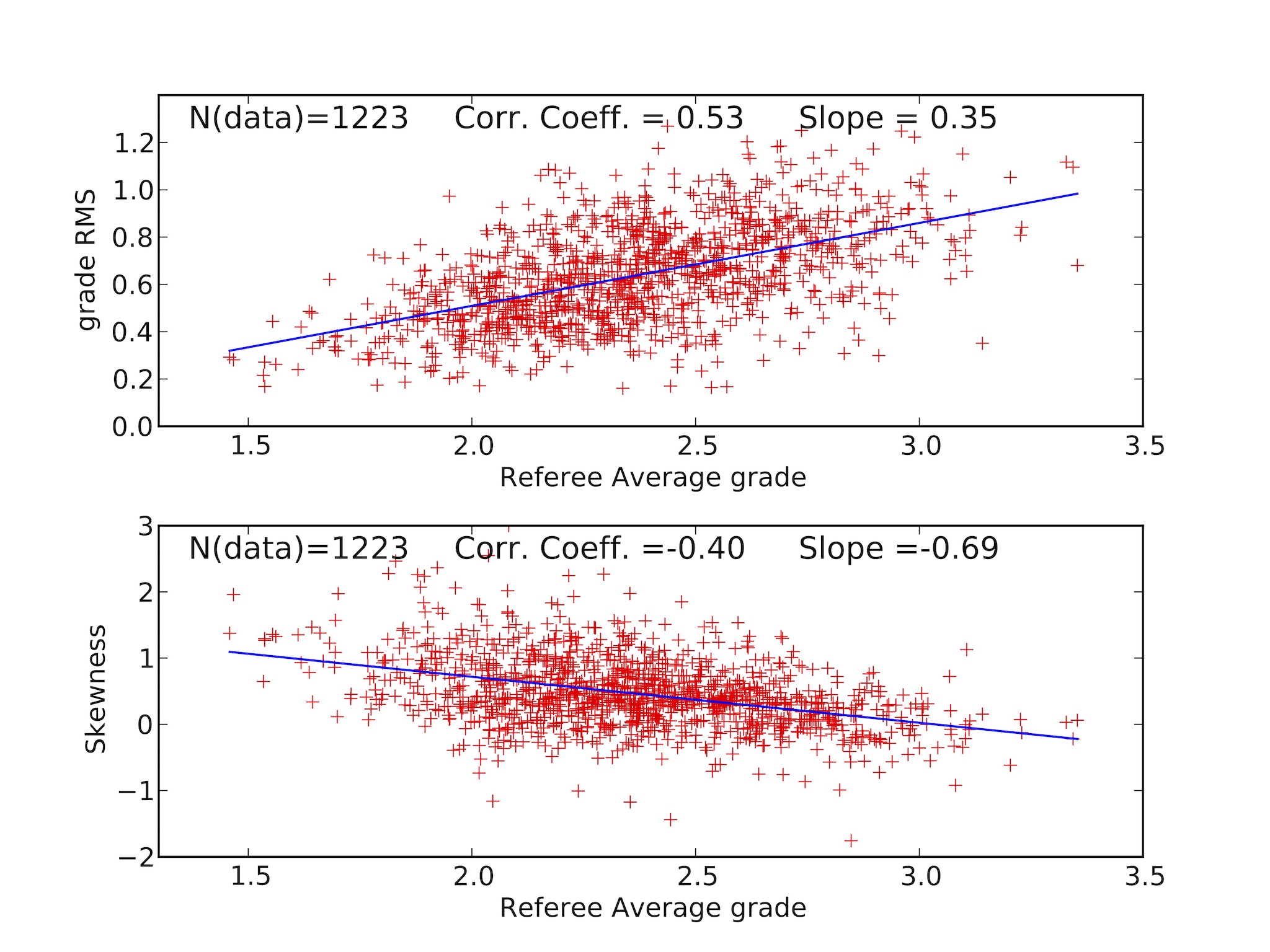}
}
\caption{\label{fig:avesig} Correlation between average grade and rms deviation (upper panel) and between average grade and Skewness (lower panel) for all referee sessions. In this and other plots, the solid blue line indicates a linear least square fit to the data. The Pearson correlation coefficient and the slope are reported on the upper left corner.}
\end{figure}
\subsection{\label{sec:sample}The sample}

The sample includes 25,469 runs (14,891 proposals) submitted by 3,110 distinct PIs ($\sim$4.8 proposals per PI, $\sim$931 proposals per semester). About 65\% of the proposals include one single run, while $\sim$85\% of the proposals have less than three runs.
On average, each proposal was reviewed by 5.6 referees, with about 95\% of the proposals reviewed by 5 or 6 referees. The proposals were reviewed by 527 distinct referees, who assigned 142,289 pre-meeting grades and 118,390 post-meeting grades. 
In the following we will distinguish between the referee and the referee session. The referee indicates the physical person and it is identified by her/his unique identifier (the referee ID). The same referee can serve one or more semesters, which we will indicate as referee sessions. These are identified with different IDs (the referee session IDs). Therefore, while a given reviewer only has one referee ID, she/he can have one ore more session IDs.

\section{General properties of the pre-meeting grades}
\label{sec:pregrades}

\subsection{Referee grade distribution}

This section presents the overall properties of the grade distribution per referee session. In the course of the study, the grade distributions of all the 1200+ referee sessions were visually inspected, revealing a great variety of behaviors. Because of this, we run a characterization of the distributions by deriving their overall statistical properties: average, root-mean-square (rms) deviation, median, semi-interquartile range (SIQR), Skewness(S) and Kurtosis\footnote{Here we use the excess Kurtosis, i.e. the Kurtosis deviation from that of a Normal distribution (3).} (K).
The distributions of some of these estimators are presented in Figure~\ref{fig:summary}. The average grade has a Normal distribution centered on 2.36 but shows a significant dispersion. The RMS deviation from the average grade has a median value of 0.63, but it also shows a wide range, from very narrow to very broad distributions.
As shown in Figure~\ref{fig:avesig} (upper panel) there is a correlation between average grade and rms deviation, in the sense that distributions centered around better grades tend to be less dispersed.

\begin{figure}
\centerline{ 
\includegraphics[width=9cm]{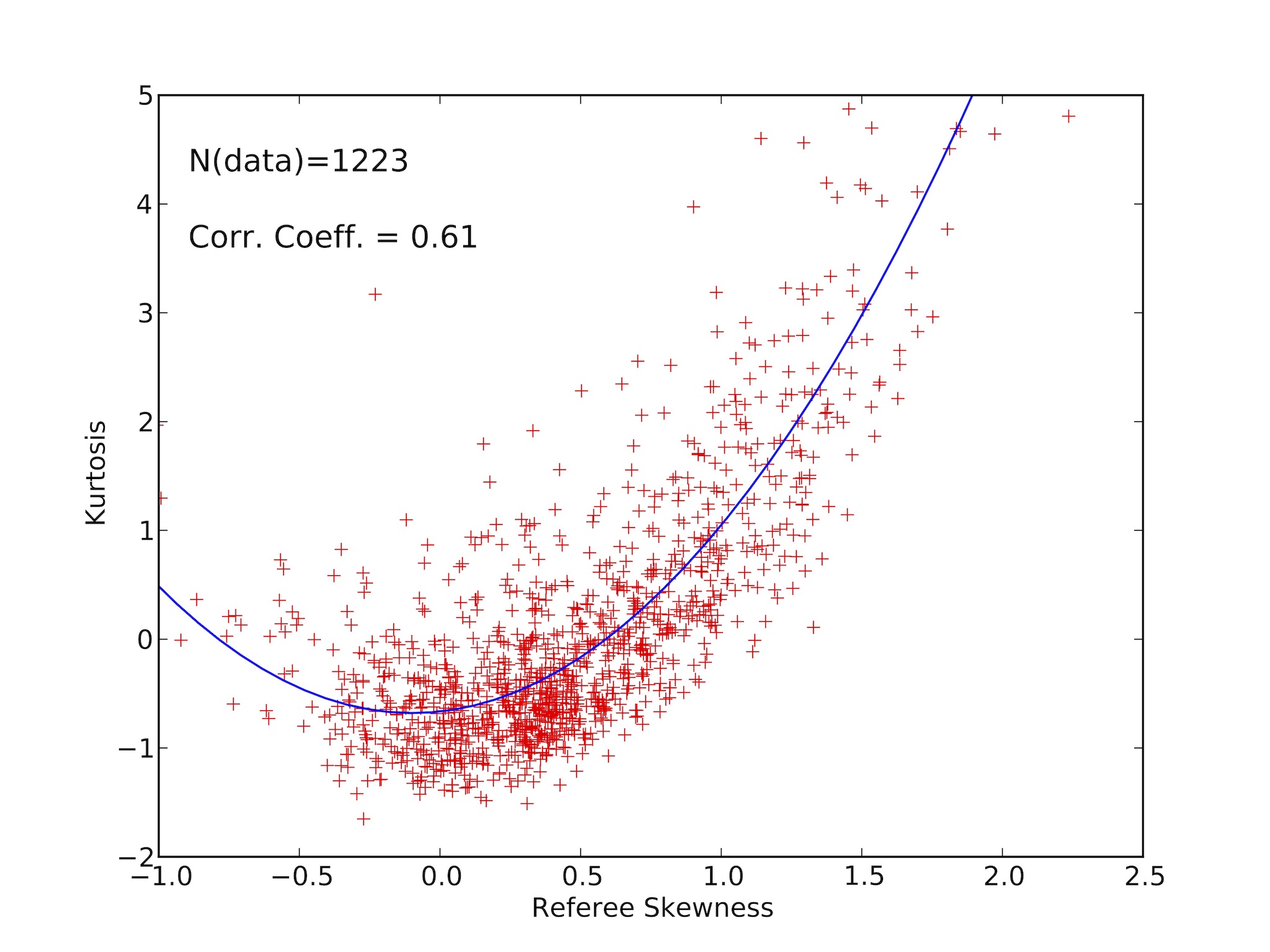}
}
\caption{\label{fig:kurskw} Correlation between Kurtosis and Skewness of referee grade distributions. The blue line traces a second order least squares fit to the data.}
\end{figure}

The single grades distributions significantly deviate from Normal. This is illustrated by the Kurtosis and Skewness distributions (see Figure~\ref{fig:summary}, lower panels). More precisely, there is a clear tendency 
for platykurtic distributions ($K<0$), which imply less extended tails. Also, there is a systematic tendency towards positive Skewness values, implying that the grade distributions tend to have an excess
at poorer grades. Another interesting correlation emerges between the Skewness and the central value: more generous distributions tend to be positively skewed, while less generous ones are negatively skewed (Figure~\ref{fig:avesig}, lower panel). This is partially explained by the fact that the referees are forced to give grades in a fixed range (1.0-5.0). For instance,  when the central value is pushed closer to high-end boundary, the distribution becomes skewed towards low-end grades.

A milder correlation is found between average grade and Kurtosis, with more generous distributions being tendentially platykurtic and less generous ones leptokurtic (extended tails). The two non-Gaussianity indicators are also found to be correlated, as illustrated in Figure~\ref{fig:kurskw}.  This is a property common to complex non-Gaussian systems (see for instance \cite{czp12}), in which K $\propto$S$^2$. The referee data appear to roughly obey a parabolic law, although the scattering is significant. 
The best fit relations are as follows:

\begin{eqnarray}
\label{eq:general}
\sigma & = & -0.19 + 0.35 \; \bar{g} \\
\nonumber
K & = & 4.03 - 1.62 \; \bar{g} \\
\nonumber
S & = & 2.11 - 0.69 \; \bar{g} \\
\nonumber
K & = & - 0.66 + 0.28 \; S + 1.43 \; S^2
\end{eqnarray}

where $\bar{g}$ is the average grade of the given referee session. In general, there are very few cases in which S$\simeq$K$\simeq$0: when S$\simeq$0, K$\sim$-0.6, while when K$\simeq$0, S$\sim\pm$0.7.

\begin{figure} 
\centerline{ 
\includegraphics[width=10cm]{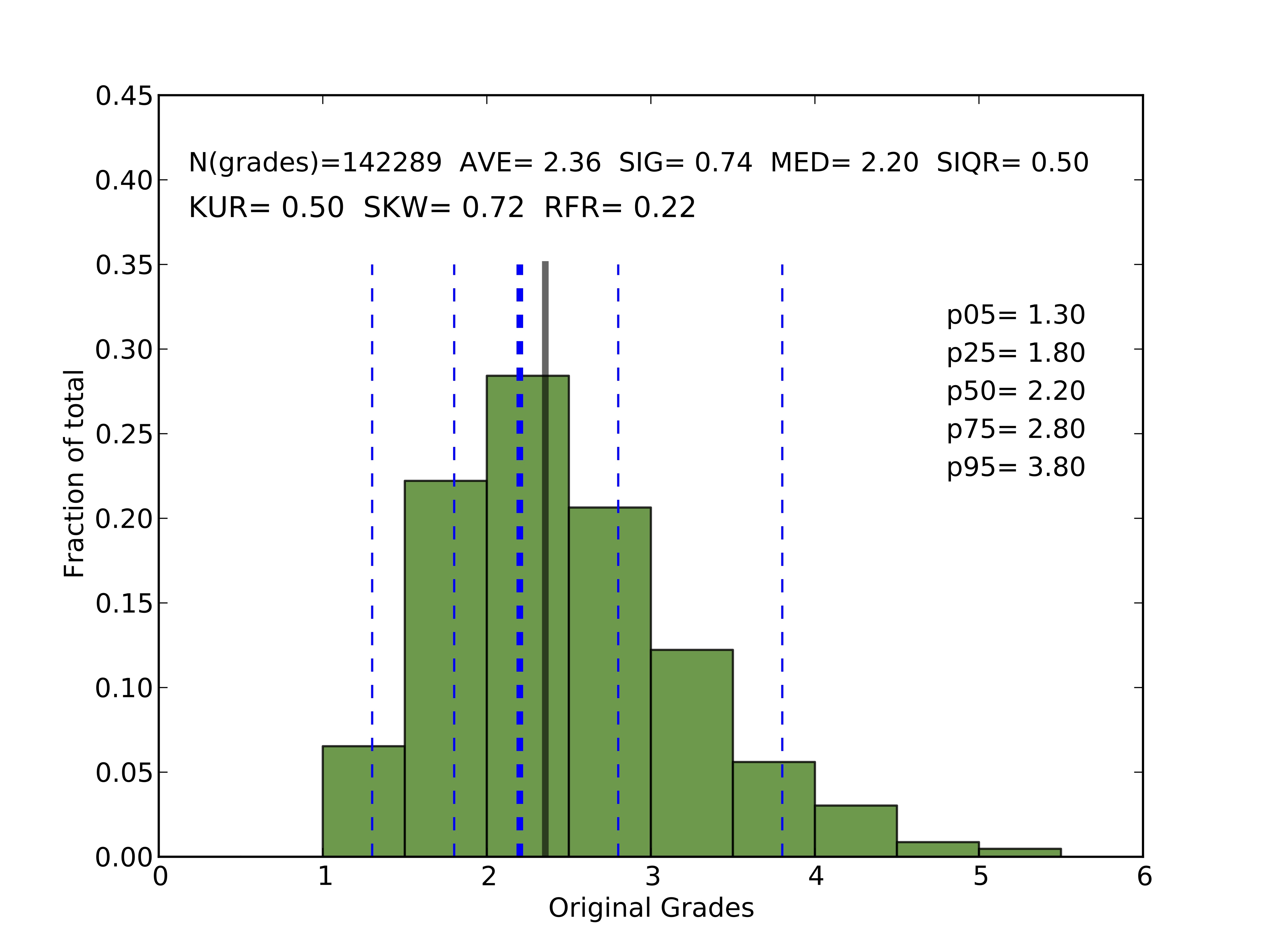}
} 
\caption{\label{fig:overall}Global distribution of pre-meeting grades. The vertical solid line
  marks the average value, while the dashed lines are placed at the
  5-th, 25-th, 50-th, 75-th and 95-th percentiles.}
\end{figure}

The sample includes 1223 referee sessions ($\sim$2.3 sessions per referee), grouped in 205 panel sessions. In about 55\% of the cases, a referee served for two semesters (typical of panel members), while about 15\% of the referees served for four semesters (typical of OPC proper members. In the reporting period range, the referees were assigned between 30 and 100 proposals per session, with a median of 70 proposals (50\% of the sessions include 63 to 77 proposals). 

\subsection{Rejection fraction}

As anticipated in Section~\ref{sec:postmeeting}, a grade larger than 3.0 implies
a deliberate run rejection. If we indicate by $f(g)$ the grade distribution function of a given
referee session, we define the rejection fraction $\rho$ as:

\begin{displaymath}
\rho = \frac{\int_{3}^{5} f(g) \; dg}{\int_{1}^{5} f(g) \; dg}
\end{displaymath}

The distribution of rejection fractions for all referee sessions in
the sample shows a marked preference for small rejection rates: $\sim$30\% of the sessions have
$\rho\leq$10\%, the median $\rho$ is about 20\%, with 95\% of the referee
sessions having $\rho\leq$53\%. As expected, there is a strong correlation between the average grade
and the rejection rate (the correlation coefficient is 0.87).

\begin{figure} 
\centerline{ 
\includegraphics[width=10cm]{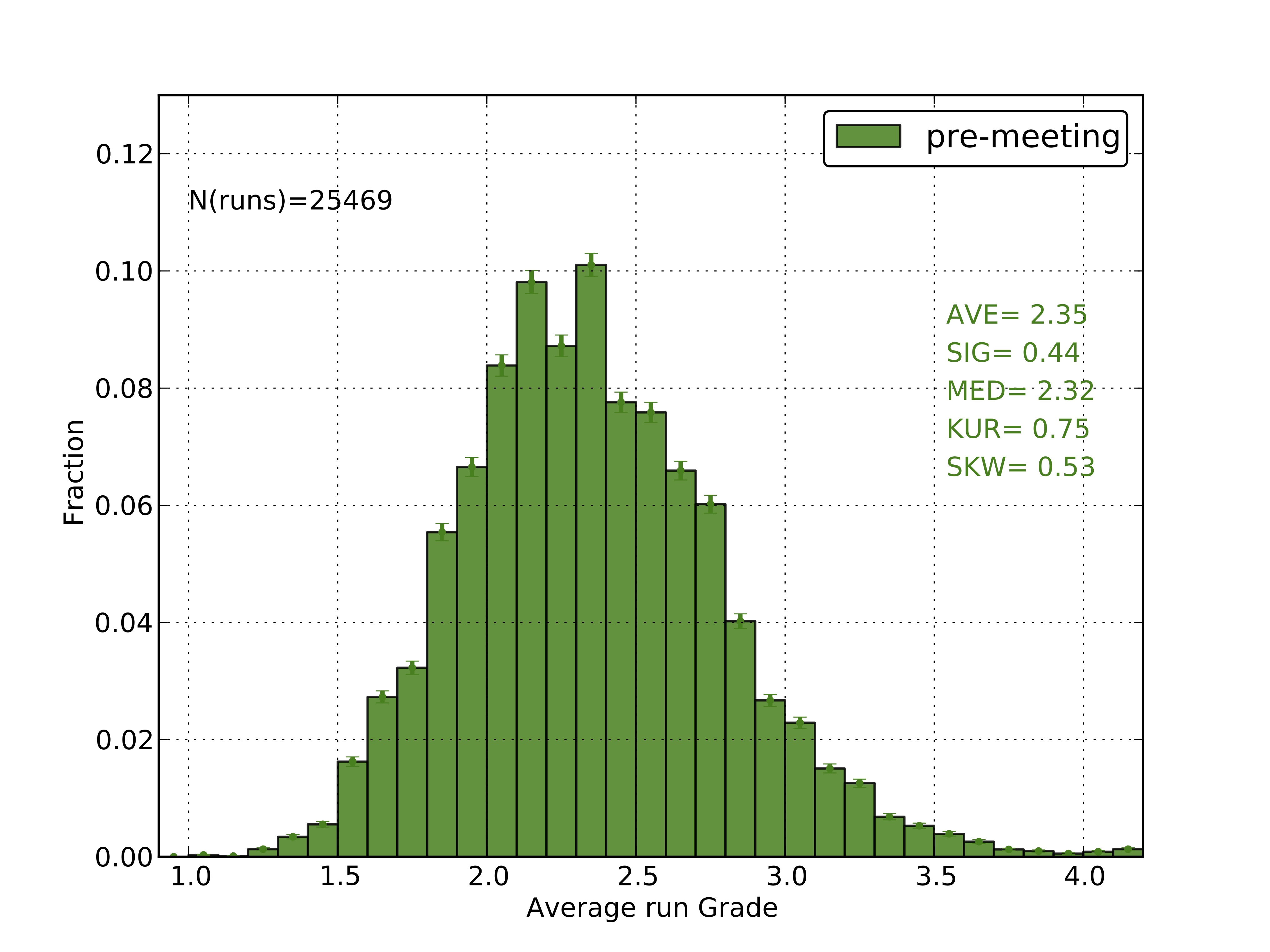}
} 
\caption{\label{fig:preruns} Pre-meeting average run grade distribution.}
\end{figure}

\subsection{Overall grades distribution}

The global pre-meeting grades distribution, obtained using all the
$\sim$140,000 grades, is presented in
Figure~\ref{fig:overall}. In this diagram the bins are kept to 0.5,
because a significant fraction of the referees give grades with that
resolution. The average grade is 2.36 (median=2.20), with an rms deviation of 0.74.
The distribution deviates from Normal, having a positive
kurtosis (0.50) and a positive skewness (0.72). The resulting
rejection fraction is 22\%.

\section{Run grades distributions}
\label{sec:rungrades}

So far we have considered the single grades independently. In this
section we will concentrate on the grades grouped by runs. 

The overall distribution, which is presented in Figure~\ref{fig:preruns}, is skewed ($S$=0.53) and has a moderate Kurtosis ($K$=0.75). The mean value is 2.35 (median=2.32), with a standard deviation of 0.44. As we will see in Paper II, this distribution is significantly modified by the discussions in the panels.

\subsection{Correlation between run average grade and standard deviation}
\label{sec:avesig}

\begin{figure} 
\centerline{ 
\includegraphics[width=10cm]{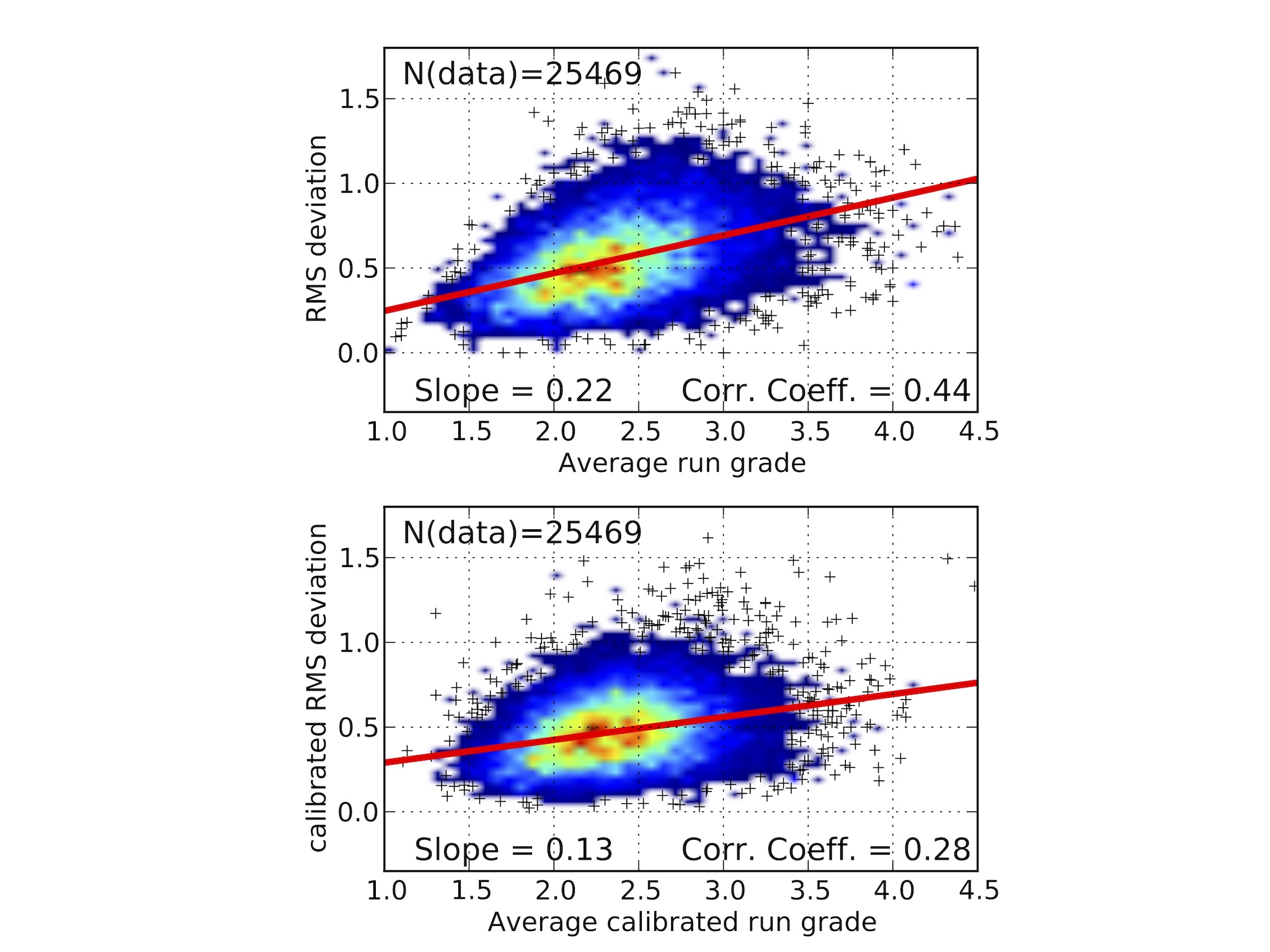}
} 
\caption{\label{fig:runavesig} Correlation between pre-meeting run average
  grade and run standard deviation for original (upper panel) and calibrated (lower panel) data. 
  The red lines trace linear best fits.}
\end{figure}

A first insight on the review process can be gained examining the behaviour 
of the standard deviation of the single grades from the average as a function
of the average itself. This is presented in Figure~\ref{fig:runavesig} (lower panel). Although 
a significant dispersion is seen, there is a tendency to have smaller deviations 
at the higher end of the grading range. 
The correlation coefficient is 0.44 (slope=0.22), and the best fit relation indicates
the rms increases by a factor $\approx$3 in the grade range
1.0-3.5. Part of this is explained in terms of numerical edge effects, caused by the fact that referees cannot give grades better than 1.0 and, 
at the same time they rarely use the full grade range at the low end of the scale (see also the discussion below).

As discussed in Section~\ref{sec:pregrades}, grade distributions characterised by 
larger average values tend to be broader. This implies that one is to expect an inherently larger scatter in the 
grades given by the various referees to the same run at the lower end of the range.

In order to quantify the amplitude of this effect, we produced a new version of the plot after applying the referee 
calibration as described in Section~\ref{sec:norm}, which reduces the effects of the systematic
differences in the various grading scales (see also Appendix~\ref{sec:calib}). The result is presented
in the lower panel of Figure~\ref{fig:runavesig}. Although the correlation factor (and the slope) decrease significantly,
there is no statistically meaningful evidence of a dispersion decrease at poorer grades.

This apparently contradict the diffuse perception that, in general, referees tend to reach a better
agreement at the high and low end of the grade range, with the mid-range being more affected by
stochastic effects. Nevertheless, as we will show in Section~\ref{sec:af}, when looking at the agreement within quartiles 
(i.e. switching from grades to ranks), the above perception is supported by statistical evidence.

In this context, it is important to bear in mind a caveat: even in the case of
complete lack of correlation between the various referees, one expects that
the dispersion decreases for average grades approaching the edges of
the allowed range. This is because, for instance, an average
grade close to 1.0 can only be produced by all grades being close to
1.0. On the contrary, average grades close to the center of the range (2.5)
can be produced by very discordant input grades. This means that the
observed decrease of the dispersion at the higher end of the
distribution is not completely a consequence of a deliberate
consensus.

\begin{figure} 
\centerline{
\includegraphics[width=10cm]{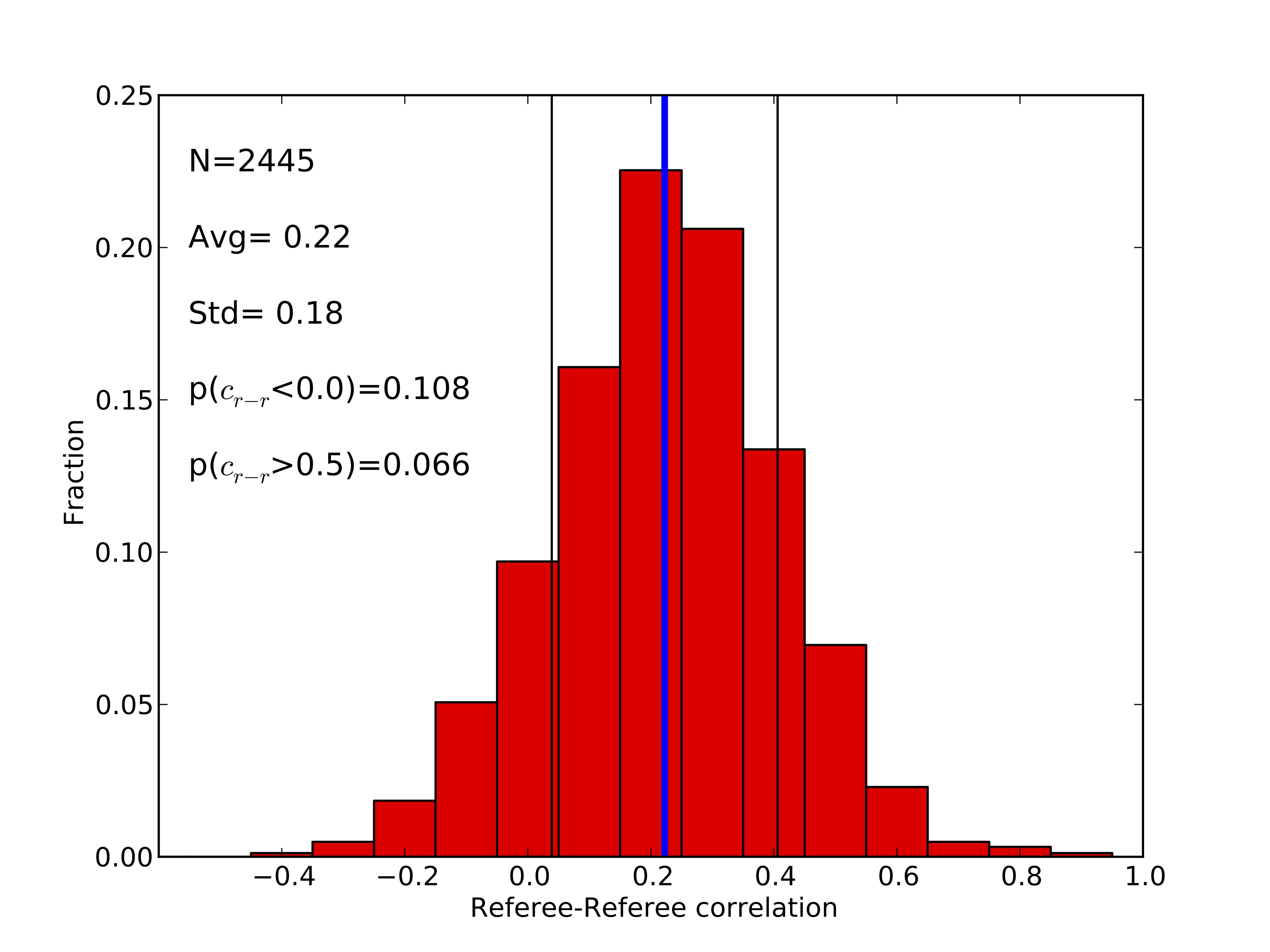}
}
\caption{\label{fig:rrcorr}The pre-meeting r-r correlation distribution. The thick vertical line indicates the average, while the two thin vertical lines mark the $\pm 1\sigma$ interval. Only runs with $N_r$=6 were included.}
\end{figure}

\section{Correlation and agreement fractions}
\label{sec:corr}

For a quantitative approach to the study of panel reproducibility, 
in this section we introduce some figures of merit that provide objective
means of  characterising the observed data, and will later serve to validate 
the model presented in Section~\ref{sec:simul}.

Some of the figures of merit introduced in this section refer to individual grades and rankings,
while others refer more generically to rank classes (e.g. quartiles). In this respect it is important
to make a distinction between selection processes with a binary outcome (accepted/rejected) and
fuzzier processes, in which the final fate of an application depends on further aspects, typically
related to scheduling constraints (see below).
Examples of the first instance are selections of talks at conferences or grant awards:
in these cases, the exact final grade (and/or rank) of accepted applications does not have 
any practical effect.

On the contrary, the individual relative ranking plays an important role in the allocation of
time in facilities like ESO, where resources at a given telescope are allotted using the
rank as a proxy to the priority assigned by the time allocation committee. For example,
two first-quartile runs (hence both formally accepted) requesting the same set of conditions 
(lunar illumination, atmospheric transparency, image quality, right ascension distribution) 
may have very different final fates, depending on their relative ranking. This dependence 
becomes more marked if the lower-ranked run has more demanding observing constraints.

\subsection{\label{sec:rrcorr}The referee-referee correlation}

Let $G_i=\{g_{i,1}, g_{i,2}, ..., g_{i,n(R)}\}$ be the set of $n(R)$ grades that were attributed by the $i$-th  referee to the
set of runs $R=\{r_1, r_2, ..., r_{n(R)}\}$ all reviewed by $N_r$ referees. We then consider the $i$-th and the $j$-th referee and we define the referee-referee (r-r, for short) correlation $c_{r-r}(i,j)$ as the Pearson correlation coefficient between the two sets $G_i$ and $G_j$. This statistical indicator provides a quantitative estimate of the overall consistency between the single referees.

Under the simplifying assumption that all runs assigned to a panel are reviewed and graded by all $N_r$ referees, one can compute $N_r (N_r - 1)/2$ distinct $c_{r-r}$ values per panel session. This can be extended to all panels and the resulting values used to construct the r-r correlation distribution. 
For the sake of simplicity, we used only runs that were reviewed by $N_r$=6 referees ($\sim$78\% of the cases)\footnote{In a small number of semesters, in order to overcame heavy loads on specific scientific categories, proposals originally assigned to a given panel had to be reviewed by members of other panels. The data corresponding to those periods were not included in the agreement analysis.}. The final selection includes 163 panel sessions (978 referee sessions, 2445 referee pairs), for a total of 96,546 grades (16,091 runs).

The resulting distribution is presented in Figure~\ref{fig:rrcorr}, The average r-r correlation is 0.22, and is accompanied by a significant standard deviation (0.18). In about 11\% of the cases the correlation is null or negative, and only in $\sim$7\% of the cases is larger than 0.5.

\begin{figure} 
\centerline{
\includegraphics[width=10cm]{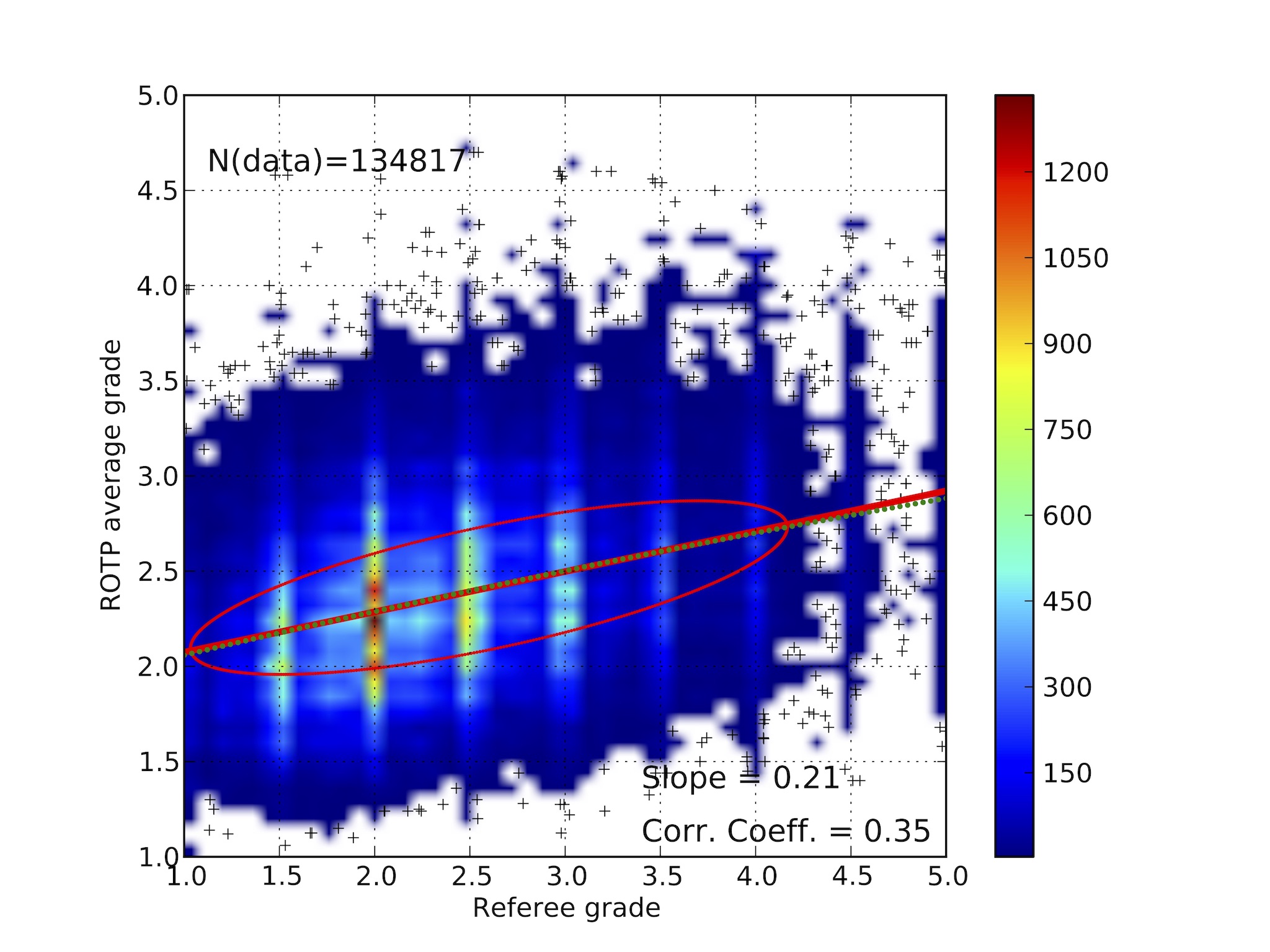}
}
\caption{\label{fig:prerrotp}The R-ROTP correlation for the pre-meeting
  grades. Only runs with 5 or 6 referees were included. The vertical density
  enhancements corresponding to 1.5, 2.0, ..., are produced by the
  referees giving grades within 0.5 bins.}
\end{figure}

\subsection{\label{sec:rotp}The referee-rest-of-the-panel correlation}

As the referee identity is not stored in the meeting phase,  the r-r correlation can only be calculated for the pre-meeting data.
To enable the comparison with the post-meeting situation (see paper II), we introduce the Referee-Rest Of the Panel correlation (R-ROTP).

For the $i$-th run and the $j$-th of the $N_r$ referees that reviewed
it, we extracted the referee grade $g_{i,j}$ and the average of the
grades given by all other referees $\tilde{g}_{i,j}$ defined as:

\begin{displaymath}
\tilde{g}_{i,j} = \frac{1}{N_r-1} \; \sum_{k=1}^{N_r-1} g_{i,k\neq j}
\end{displaymath}

The pairs $(g_{i,j}, \tilde{g}_{i,j})$ are then considered as points
on a two-dimensional diagram and the correlation coefficient is
computed. The pre-meeting R-ROTP correlation is presented in
Figure~\ref{fig:prerrotp}, only for runs with $N_r\geq$5 (which cover
about 85\% of the cases). 

The correlation coefficient is 0.35, indicating a weak, but non-null
correlation between the grades. 

\begin{figure} 
\centerline{
\includegraphics[width=10cm]{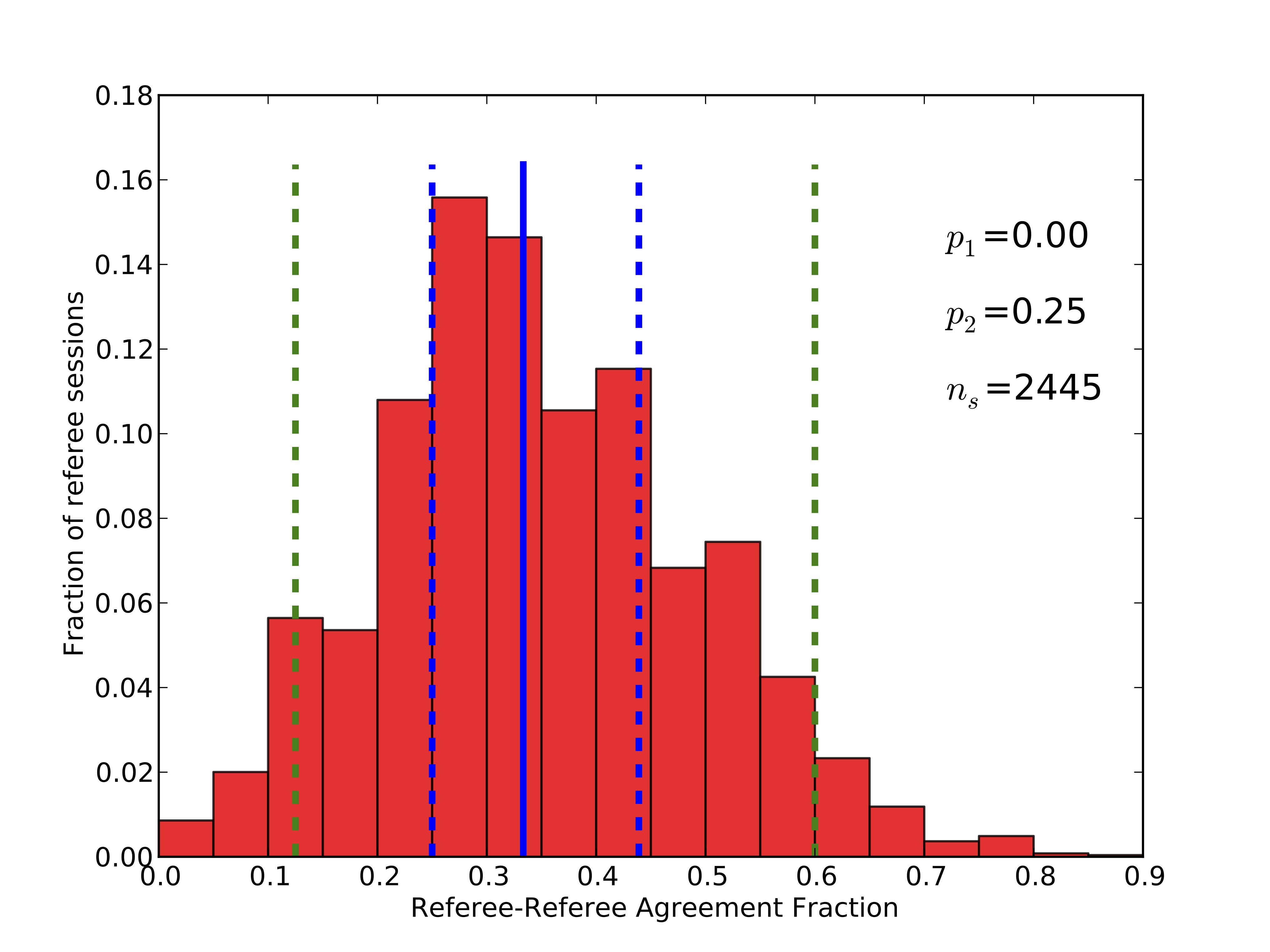}
}
\caption{\label{fig:preagrfrac}Pre-meeting first quartile r-r agreement fraction. The solid vertical line
marks the average value. The dashed lines indicate the 5-th, 25-th, 75-th and 95-th percentiles.}
\end{figure}

\subsection{The agreement fractions}
\label{sec:agfr}

A different figure of merit for characterising the level of correlation between the grades given by the various referees is what we will indicate as the agreement fraction (see Appendix~\ref{sec:appa}). For the sake of simplicity (but without any significant loss of statistical significance), we have used the same sub-sample as in Section~\ref{sec:rrcorr} (only runs with $N_r$=6). These estimators implicitly eliminate the grades calibration problem (see Appendix~\ref{sec:calib}), as they are practically based on the ranking rather than on the grading (see however footnote~\ref{foot:random} for the caveats one should bear in mind when deriving the fractions from graded rather than ranked runs).

For our purposes we introduce different types of agreement fractions.

\subsubsection{The referee-referee agreement fraction}
\label{sec:af}

The referee-referee (r-r) agreement fraction $f_{r-r}$ (see Appendix~\ref{sec:ira}) is the fraction of runs graded by referee $i$  within a certain percentile interval ($p_1 \leq p \leq p_2$) of the grade distribution\footnote{\label{foot:random}The grade distributions were computed using all the runs assigned to a given referee, also those that were not graded by all $N_r$ referees. In order to overcome the problem associated with the grade binning and percentile calculations, the grades were randomised adding a uniform noise with a maximum semi-amplitude of 0.01. Multiple runs of the calculations show that the results depend very weakly on the specific random realisation. This is equivalent to assuming that runs given the same grade would get a random relative rank.}, that were also selected by referee $j$, member of the same panel. For each panel session, this leads to $N_r (N_r-1)$ distinct values of $f_{r-r}$. Their distribution is shown in Figure~\ref{fig:preagrfrac} for the first quartile ($p_1=0$, $p_2$=0.25).

The average agreement fraction is 34\%, with a standard deviation of 14\% (the 95-th percentile is 59\%). The r-r agreement is less than 50\% in 86\% of the cases.
It is worth noticing that the average r-r agreement fraction expected for complete uncorrelation between the grades given by two referees for their respective first quartiles is 25\%. This can be expressed using the Cohen's kappa coefficient ($k$) \citep{cohen}, which takes into account the chance agreement ($k$=0 for a purely random agreement, $k$=1 for perfect agreement). The first quartile r-r agreement fraction quoted above corresponds to $k$=0.12.

When discussing the sub-panel agreement (Section~\ref{sec:subpanel}) and simulations (Section~\ref{sec:simpre}), we will extend the concept to panels, by introducing the panel-panel (p-p) agreement fraction $f_{p-p}$. It is worth noticing that, since the agreement fraction is based on the rankings (and not on the grades), it is independent from the actual shape of the single grade distributions.

\subsubsection{Referee-referee quartile agreement matrix}

One can extend the above concept to derive what we call the referee-referee quartile agreement matrix. Its elements $f^{k,l}_{r-r}$ are the average r-r agreement fractions between the runs ranked in quartile $k$ by referee $i$ and the runs ranked in quartile $l$ by referee $j$. The diagonal elements ($k=l$) coincide with the average r-r agreement fraction discussed above. The matrix is presented in Table~\ref{tab:ira}. Interestingly, on average, about 17\% of the runs ranked in the first quartile by a referee are ranked in the fourth quartile by any other panel member, while about 25\% of the runs placed in a given quartile by a referee are ranked in one of the adjacent quartiles by another referee. Obviously, in a purely random process all the matrix elements would be equal to 0.25, while in a completely correlated case, all non-diagonal elements would be null, and diagonal elements would be 1.0.

\begin{table}
\caption{\label{tab:ira}Referee-Referee Quartile Agreement Matrix}
\tabcolsep 5.0mm
\begin{tabular}{c|cccc}
\hline
\hline
 first ref.  & \multicolumn{4}{c}{second referee quartile}\\
 quartile & 1 & 2 & 3 & 4\\
\hline
1 & 0.34 &  0.26 & 0.22 &  0.18\\
2 & 0.26 &  0.26 & 0.25 &  0.23\\
3 & 0.22 &  0.25 & 0.27 &  0.27\\
4 & 0.17 &  0.21 & 0.26 &  0.36\\
\hline
\end{tabular}
\end{table}

An interesting aspect is that the r-r agreement in the first and fourth quartiles ($f^{1,1}$, $f^{4,4}$) is larger than in the two central quartiles ($f^{2,2}$, $f^{3,3}$), where it is only marginally above the random limit (see Table~\ref{tab:ira}; $k$=0.01--0.03). As we will see in the simulations (Section~\ref{sec:simul}), this difference is related to the fact that the same uncertainty in the grade scale produces percentile variations that are larger as one approaches the central value (i.e. in the two central quartiles).

\subsubsection{Referee-majority agreement fraction}

The referee-majority (r-m) agreement fraction $f_{r-m}$ is defined in Appendix~\ref{sec:rm}. Despite its apparent complexity, $f_{r-m}$ is simply the fraction of runs graded within a certain percentile interval ($p_1 \leq p \leq p_2$) by the given referee that were graded in the same interval by at least 50\%+1 of the panel (4 members in our specific case). It therefore expresses the fraction of occurrences in which any referee agrees with the majority.

This produces a number of values equal to the number of referee sessions. For the first quartile and for runs that have $N_r$=6, the average r-m agreement fraction is 23\% and is accompanied by a large spread (the standard deviation is 14\%). The agreement fraction is less than 50\% in about 96\% of the cases. In other words, in most of the occurrences the number of runs for which there is agreement (in the above defined sense), is smaller than the number of runs for which this agreement is not reached. As one can show with simple Monte-Carlo simulations, in case of complete uncorrelation $f^{1,1}_{r-m}\sim$10\%.

In this context one can also compute the overall majority agreement fraction. For any given quartile, this is defined as the ratio between the number of runs that were ranked in that quartile by the majority of the panel and the number of runs expected in each quartile (a quarter of the total number of runs). In simpler words, this is the fraction of runs for which there was "democratic consensus". For the $N_r$=6 case, this is 39\% for the first quartile, decreasing to 18\% in the central quartiles. The simulations show that, for a completely random process, the fraction is $\sim$15\%.

\subsection{Sub-panel agreement fraction}
\label{sec:subpanel}

A direct, quantitative measurement of the agreement between two distinct panels requires the availability of a statistically significant sample of proposals assigned to both of them. Although our data do not include such a case, it is nevertheless possible to bootstrap the existing data to get some direct insight\footnote{This is not possible for the post-meeting phase, in which the available data are affected by the discussion, and it is therefore impossible to reconstruct what the outcome would be if the panel had split into two sub-panels before the discussion took place. See Paper II.}. For this purpose, we used only pre-meeting cases with $N_r$=6 and, for each panel session, we computed the quartile agreement fractions for the 10 non-intersecting pairs of sub-panels including 3 members each. This provides 1630 different quartile agreement matrices, from which the average and standard deviation matrices were finally computed. The result is presented in Table~\ref{tab:boot}. The first quartile agreement is about 43\%, while for the second and third quartile this is only about 30\%, i.e. slightly above the 25\% value of a purely random process. The typical standard deviation of the distributions of single fractions ranges from 0.09 to 0.15. 

Using the above bootstrap procedure we have checked the agreement at the two extremes of the ranking, computing $f_{p-p}$ in the first and last decile. We have done this because, in our experience, there is a diffuse perception about referees being able to almost unanimously identify the top (and bottom) cases. The data yield 0.22 and 0.30 for the average fractions in the first and the last decile, respectively, while the values are close to the random limit (0.10) in all other cases. Therefore, although this finding is in line with the enhanced agreement observed in the first and fourth quartiles, it also shows that the agreement at the very extremes of the ranking is not larger. Therefore, there is no indication for the existence of a common sub-set of proposals that would be placed at the top of their rank by any number of independent panels. This is in line with the findings published by \cite{pier}. In their study on National Institutes of Health grant applications, they report that 'reviewers do no reliably differentiate between good and excellent grant applications'.

\section{An heuristic approach}
\label{sec:model}

In this section we discuss an heuristic approach, with the final aim of making 
predictions on the reproducibility of a panel ranking in the pre-meetings phase.
The model is based on the following working hypothesis, which we will indicate
as the True Grade Hypothesis (TGH):

\vspace{2mm}

{\it For any given proposal a true grade does exist. The true grade can be determined as the
average of the grades given by a very large number of referees.}

\vspace{2mm}

\begin{table}
\caption{\label{tab:boot}Bootstrapped sub-panel Quartile Agreement Matrix ($N_r$=3).}
\tabcolsep 5.0mm
\begin{tabular}{c|cccc}
\hline
\hline
 first s.-p. & \multicolumn{4}{c}{second sub-panel quartile}\\
 quartile & 1 & 2 & 3 & 4\\
\hline
1 & 0.43 &  0.28 & 0.19 &  0.10\\
2 & 0.25 &  0.30 & 0.26 &  0.20\\
3 & 0.18 &  0.25 & 0.29 &  0.29\\
4 & 0.09 &  0.18 & 0.27 &  0.46\\
\hline
\end{tabular}
\end{table}

The first part of hypothesis is obviously debatable, as it implicitly
assumes that an absolute and infinitely objective scientific value can be
attached to a given science case. It does not take into account, for
instance, that a proposal is to be considered within a certain context
and cannot be judged in vacuum.  Most likely, a proposal requesting time to measure the positions of stars 
during a total solar eclipse would have been ranked very highly in 1916, 
but nowadays it would probably score rather poorly. The science case
is still very  valuable in absolute terms, but is deprived of almost any interest 
in the context of modern physics.

The second part of the hypothesis is also subject to criticism,
because it implicitly assumes that referees behave like objective
measurement instruments. This is most likely not the case.  For
instance, although the referees are proactively instructed to focus
their judgment on the mere scientific merits, it is unavoidable that
they (consciously or unconsciously) take into account (or are
influenced by) other aspects. Among these are the previous history of
the proposing team, its productivity, its public visibility, the
inclusions of certain individuals, personal preferences for specific 
topics, and so on.

In the following we will present a simple model  based on the TGH. We separate the pre-meeting grading process into two components: subjective and systematic.

\subsection{Subjective component}

The subjective part can be described as follows:

\begin{enumerate}
\item For any given proposal a true grade $\gamma$ exists;
\item The true grade $\gamma$ obeys to a distribution $\tau(\gamma)$;
\item The subjective part behaves as an additive noise $\Delta \gamma$ characterised by a distribution $e(\Delta \gamma)$. This leads to the measured value $g\prime=\gamma+\Delta \gamma$.
\item The grade distribution after the measurements is the convolution 
$t(g\prime)= \tau(\gamma) \ast e(\Delta \gamma)$.
\end{enumerate}

We note that the noise function describes the overall fuzziness of the process, without specifying where the source of the uncertainty is. 

\begin{figure} 
\centerline{
\includegraphics[width=9cm]{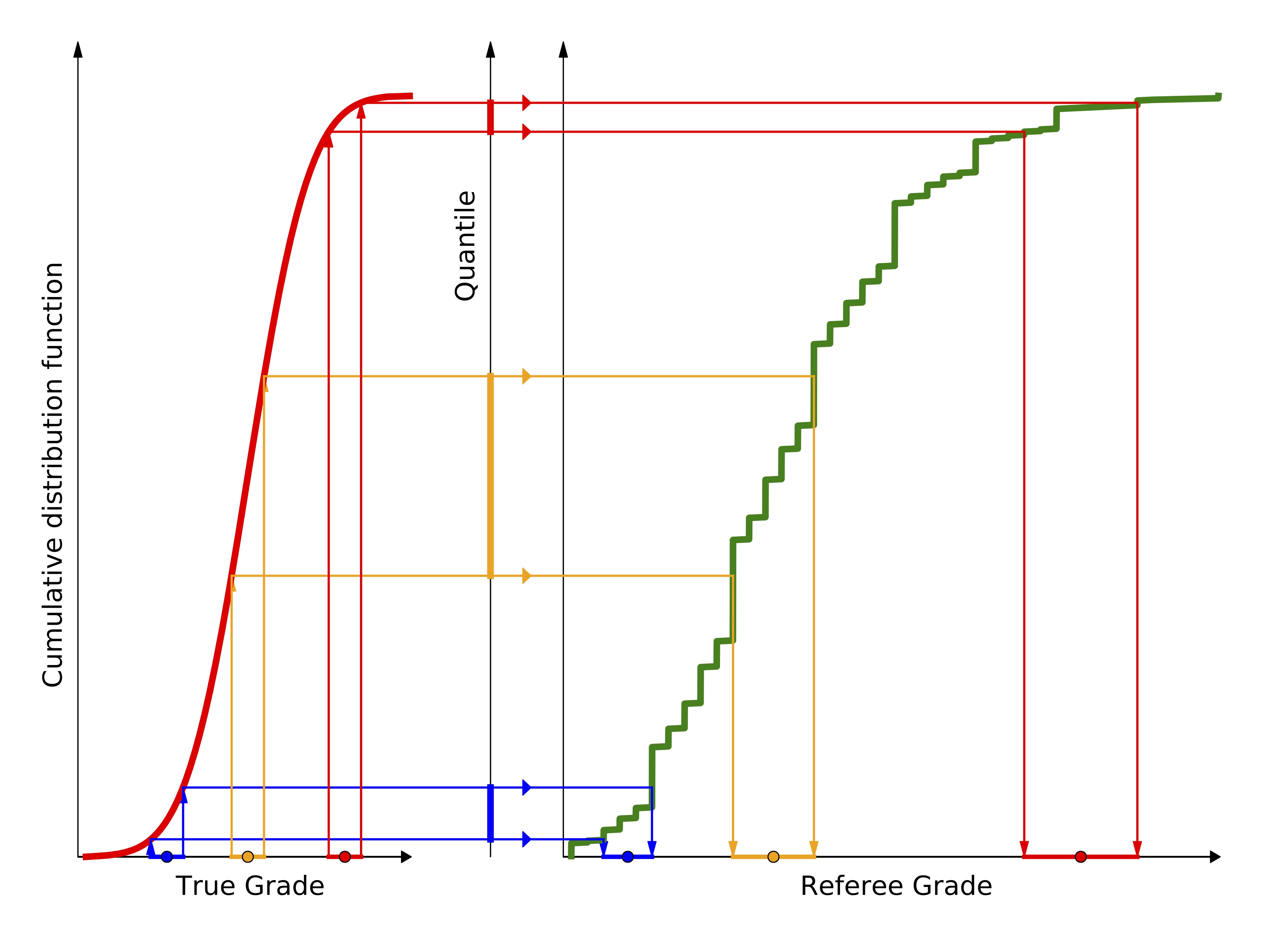}
}
\caption{\label{fig:process}Schematic representation of the true to referee grade transformation process. For illustration purposes we have used the average cumulative function derived from all pre-meeting referee sessions in the sample (green curve).  The vertical jumps correspond to the 0.5 grade bins used by many reviewers.}
\end{figure}

\subsection{Systematic component}

As we have seen in Section~\ref{sec:pregrades}, there is a great variability in the properties of the grade distributions. This is modelled through the systematic process which, by construction, does not introduce any random component:

\begin{enumerate}
\item Each referee is characterised by a distribution $r(g)$, with a
  dispersion (describing the dynamical range of grades) and a central value (related to the offset of this range with respect to the true grade distribution). This generalizes the concept of "personal equation" \citep{schaffer}.
\item The "blurred" grade $g\prime$ is converted to a grade $g$ in the referee scale by a transformation that conserves its quantile (and hence its ranking). If $T(g\prime)$ and $R(g)$ are the cumulative distribution functions of the probability density functions $t(g\prime)$ and $r(g)$, the transformation can be written as: $g = R^{-1}[T(g\prime)]$, where $R^{-1}$ is the quantile function, i.e. the inverse of the cumulative distribution function of $r(g)$.
\end{enumerate}

With these settings, the pre-meeting grading process can be described as follows:

\begin{equation}
\label{eq:model}
g = R^{-1}[T(g\prime)] \equiv R^{-1}[T(\gamma + \Delta \gamma)]
\end{equation}

We note that with this formulation the shape of the observed referee distribution $r(g)$ is independent from $\tau(\gamma)$. Although the simulations show that the results are indistinguishable, one can in principle apply the blurring after transforming the true grades $\gamma$ to the referee scale. We preferred the choice described above, as it allows us to constrain the shape of $r(g)$ directly from the observed distributions.

In the following we will indicate with $\gamma_0$ and $\sigma_\gamma$ the average and the dispersion of the true grade distribution, and with $\sigma_e$ the dispersion of the subjectivity distribution $e(\Delta \gamma)$, which in principle can depend on $\gamma$. We then indicate with $\alpha$ the ratio:

\begin{equation}
\label{eq:alpha}
\alpha = \frac{\sigma_e}{\sigma_\gamma} 
\end{equation}

which sets the level of 'confusion' in the pre-meeting process.

A schematic representation of the process is given in Fig.~\ref{fig:process}, in which for $R(g)$ we used the average cumulative distribution function derived from the 1223 referee sessions. For illustrative purposes we indicated the transformation of three different true grades, affected by the same uncertainty (coloured bars on the left x-axis), which produces grade-dependent uncertainties on the referee grades (coloured bars on the right x-axis). The exact dependency is dictated by the shape of $R(g)$, but in general the uncertainty grows at poorer grades.  On the other hand, the uncertainty on the quantile (and hence on the ranking) depends only on $\alpha$, and not on the distribution characterising the referee (see the vertical coloured bars on the quantile axis of Figure~\ref{fig:process}).

\section{Model validation}
\label{sec:validation}

\subsection{Modelling single referees}

In a first exploratory approach, we used truncated Gaussians for $r(g)$. These have the advantage of providing grades within the required range (1-5), and well describe the shapes observed in the real data with only two parameters (the central value and the dispersion). For instance, a truncated Gaussian with an underlying $\sigma\geq$2 gives a flat distribution, while pushing the central value close to the two edges naturally produces positively or negatively skewed distributions (see Appendix~\ref{sec:truncated}). The simulations show that truncated Gaussians provide a satisfactory description of the observed data, especially in reproducing the average values. However, the resulting synthetic distributions of the statistical indicators introduced above (r-r correlations, agreement fractions, etc.) have tails that are less extended than those seen in our sample, as they provide smooth distributions, without the observed outliers.

Because of this and given the wealth of available data, we decided to follow a different path: the grades are generated directly using the observed cumulative distribution functions. Given the size of the sample of real referee sessions, this provides a sufficiently dense sampling of real-life behaviour.

\begin{figure} 
\centerline{
\includegraphics[width=10cm]{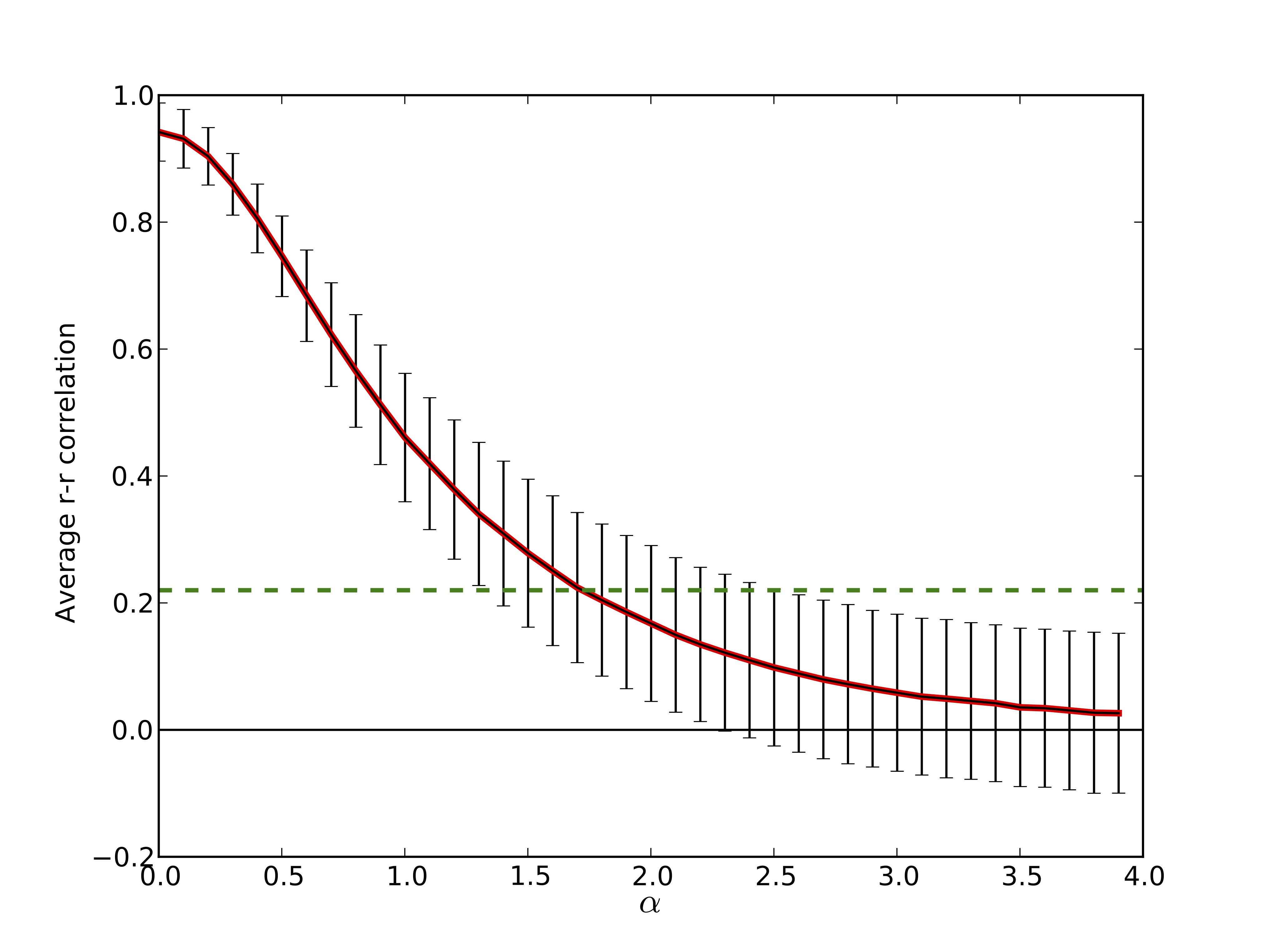}
}
\caption{\label{fig:simrrcorr}The average referee-referee correlation as a function of $\alpha$. The horizontal dashed line marks the
measured value (0.22). The error-bars indicate the standard deviation of the simulated r-r correlation distribution for the given $\alpha$.}
\end{figure}

For each observed referee session we derived the normalized cumulative distribution function $R(g)$. We then computed the quantile functions $R^{-1}$ to be used in Eq.~\ref{eq:model} by numerical inversion, and stored the results in a look-up table. When a synthetic panel is generated, a set of $N_r$ distinct quantile functions are randomly selected from the $\sim$1200 entries in the look-up table, and the corresponding grades derived via linear interpolation.

For $\tau(\gamma)$ and $e(\Delta \gamma)$ we assumed Normal distributions centred on $\gamma_0$=0 and characterised by dispersions $\sigma_\gamma$ and $\sigma_e$, which are related to each other via Eq.~\ref{eq:alpha}.

The $\alpha$ ratio is the only free parameter in our model (see Eq.~\ref{eq:alpha}). In this section we will constrain its value using the pre-meeting data, and verify that the statistical indicators that we introduced are properly matched.

In all simulations we assigned 72 applications to each referee, i.e. the real average for ESO panels.

\begin{figure} 
\centerline{
\includegraphics[width=10cm]{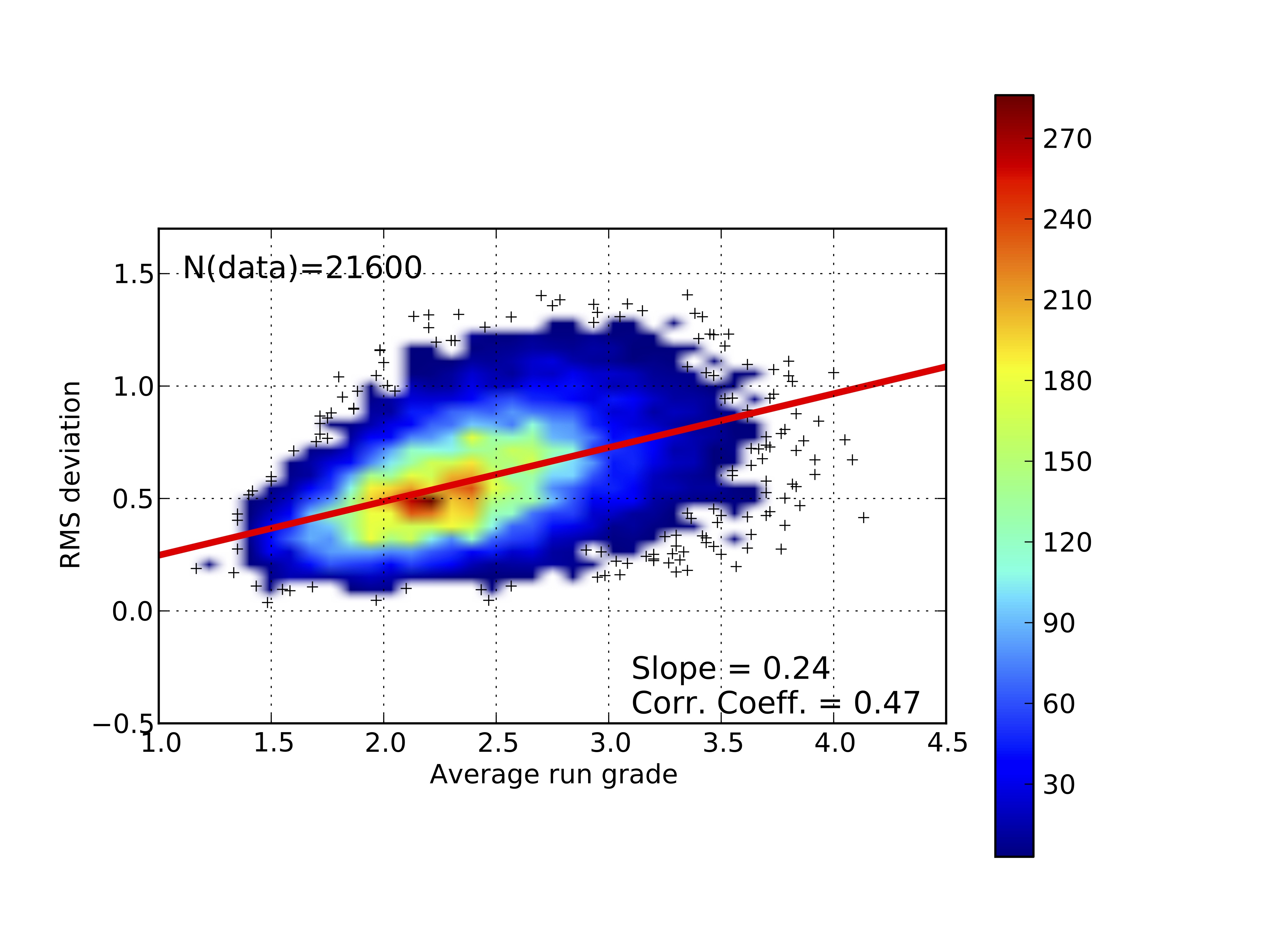}
}
\caption{\label{fig:simrunavesig} Correlation between simulated pre-meeting run average grade and run standard deviation.}
\end{figure}

\subsection{Pre-meeting r-r correlation}

For determining $\alpha$ we simulated a large set of runs reviewed by random pairs of referees, and determined the distribution of the r-r correlation (see Section~\ref{sec:rrcorr}). With this set-up we derived the average $c_{r-r}$,  and varied the value of $\alpha$ until a best match to the data  was reached. 

The dependence is illustrated in Figure~\ref{fig:simrrcorr}. The fraction rapidly decreases from $\approx$100\% (attained for $\alpha$=0), and asymptotically reaches 0\% for $\alpha>4$, which corresponds to a complete un-correlation. The measured average correlation (0.22, see Section~\ref{sec:rrcorr}) is attained for $\alpha$=1.7, which indicates that the uncertainty of the process is significantly larger than the intrinsic dispersion of the true grades.

\subsection{Pre-meeting referee-referee agreement}

A first validation test for the proposed model was run computing the r-r quartile agreement matrix for $\alpha$=1.7.

The resulting matrix is presented in Table~\ref{tab:simafm}. This is remarkably similar to the observed matrix (Table~\ref{tab:ira}), to the level that it reproduces the slight asymmetry observed for the fourth-quartile, generated by the fact that the observed average distribution is not centred on the centre of the allowed range (3.0).
An important aspect worth being emphasized is the increased agreement fractions for the first and fourth quartile, observed both in the data and in the simulations. In the context of the adopted model, this is a natural consequence of keeping  the confusion level (regulated by $\alpha$) constant. More precisely, this is caused by the fact that the function that relates a grade to its percentile is not linear: the same grade fluctuation produces changes on the percentile which depend on the grade itself. In particular, the same grade change in the first and fourth quartile produce rank changes that are smaller than in the central quartiles (see also Appendix~\ref{sec:calib}).

\subsection{Pre-meeting correlations}

The next step is the matching of the global statistical properties of the referee grade distributions after fixing $\alpha=1.7$. For this purpose we generated a large number of referee sessions and analysed the properties of the resulting distributions in the same way as for the real data (Section~\ref{sec:pregrades}). The distributions and the correlations between the various statistical estimators  are all very well reproduced. An example of the level of matching is shown in Figure~\ref{fig:simrunavesig}, which reproduces very closely the observed behavior, especially when comparing to the real data with $N_r$=6 (in this case the slope and the correlation coefficient are practically identical to those resulting from the simulations).

\begin{table}
\caption{\label{tab:simafm}Simulated r-r Quartile Agreement Matrix ($\alpha$=1.7).}
\tabcolsep 5.0mm
\begin{tabular}{c|cccc}
\hline
\hline
 first ref.  & \multicolumn{4}{c}{second referee quartile}\\
 quartile & 1 & 2 & 3 & 4\\
\hline
1 & 0.33 &  0.26 & 0.22 &  0.17\\
2 & 0.26 &  0.25 & 0.24 &  0.22\\
3 & 0.22 &  0.24 & 0.25 &  0.27\\
4 & 0.16 &  0.21 & 0.26 &  0.37\\
\hline
\end{tabular}
\end{table}

It is worth noting that this weak dependence is reproduced without the need for making the blurring effect dependent on the grade (by using a variable $\alpha$). This implies that the observed dependence is not due to an increased intrinsic fuzziness in the process at poorer grades, but rather to the observed fact that the distributions of single referees tend to become more dispersed for larger central values.

\section{Simulations}
\label{sec:simul}

Having ascertained that the model provides a satisfactory match to the real sample, it is now possible to use it for predicting the reproducibility of the review process, which is one of the core matters we set out to tackle.

Although in the specific case of ESO what counts is the end result of the panel discussions (see Paper II), for the sake of generality we discuss the pre-meeting phase as well, as there are many common instances in which the final ranking is fully based on independent (at home) referee evaluation, without having the reviewers meet and discuss in person. The effects of the panel discussion will be addressed in great detail in Paper II.

\subsection{Referee and panel precision}

A first simple set of simulations allows to estimate the expected precision of a single referee, here meant as the typical deviation from the true grade. This is done 'submitting' applications characterised by the same true grade to the same referee, and repeating the Monte-Carlo simulation for  different input grades and referees. This allows one to compute the average rms dispersion, which turns out to be $\bar{\sigma}$=0.57 (in the 1 to 5 scale adopted at ESO). This number gives an immediate idea of the 'signal-to-noise' that one is to expect on the rating produced by one single, independent reviewer. 
The precision obviously improves increasing the number of referees: as the simulations show, $\bar{\sigma}\propto \sqrt{N}$. For the typical ESO setup ($N_r$=6) this yields $\bar{\sigma}$=0.23. In Appendix~\ref{sec:calib} we show how this propagates into the rank uncertainty.

\begin{table}
\caption{\label{tab:simppaf}Simulated pre-meeting p-p quartile agreement matrix}
\tabcolsep 5.0mm
\begin{tabular}{c|cccc}
\hline
\hline
   panel \#1& \multicolumn{4}{c}{panel \#2 quartile}\\
quartile  & 1 & 2 & 3 & 4\\
\hline
1 & 0.55 &  0.27 & 0.13 &  0.04\\
2 & 0.27 &  0.33 & 0.27 &  0.13\\
3 & 0.13 &  0.27 & 0.33 &  0.28\\
4 & 0.04 &  0.13 & 0.28 &  0.55\\
\hline
\end{tabular}
\end{table}

\subsection{Panel reproducibility}
\label{sec:simpre}

The reproducibility of two independent pre-meeting panels reviewing the same set of proposals can be quantified through the panel-panel quartile agreement matrix elements $f^{k,l}_{p-p}$, similarly to what we did for the single referees (see Section~\ref{sec:af}). The average agreement matrix is presented in Table~\ref{tab:simppaf} for $N_r$=6.  Two panels agree on  the top quartile in more than half of the cases ($f_{p-p}^{1,1}$=55.0$\pm$16.7\%, 95\% confidence level). The agreement decreases in the central quartiles ($\sim$33\%), to exceed again 50\% in the fourth quartile. The corresponding values for Cohen's kappa are 0.40 and 0.11, respectively.

\begin{figure} 
\centerline{
\includegraphics[width=10cm]{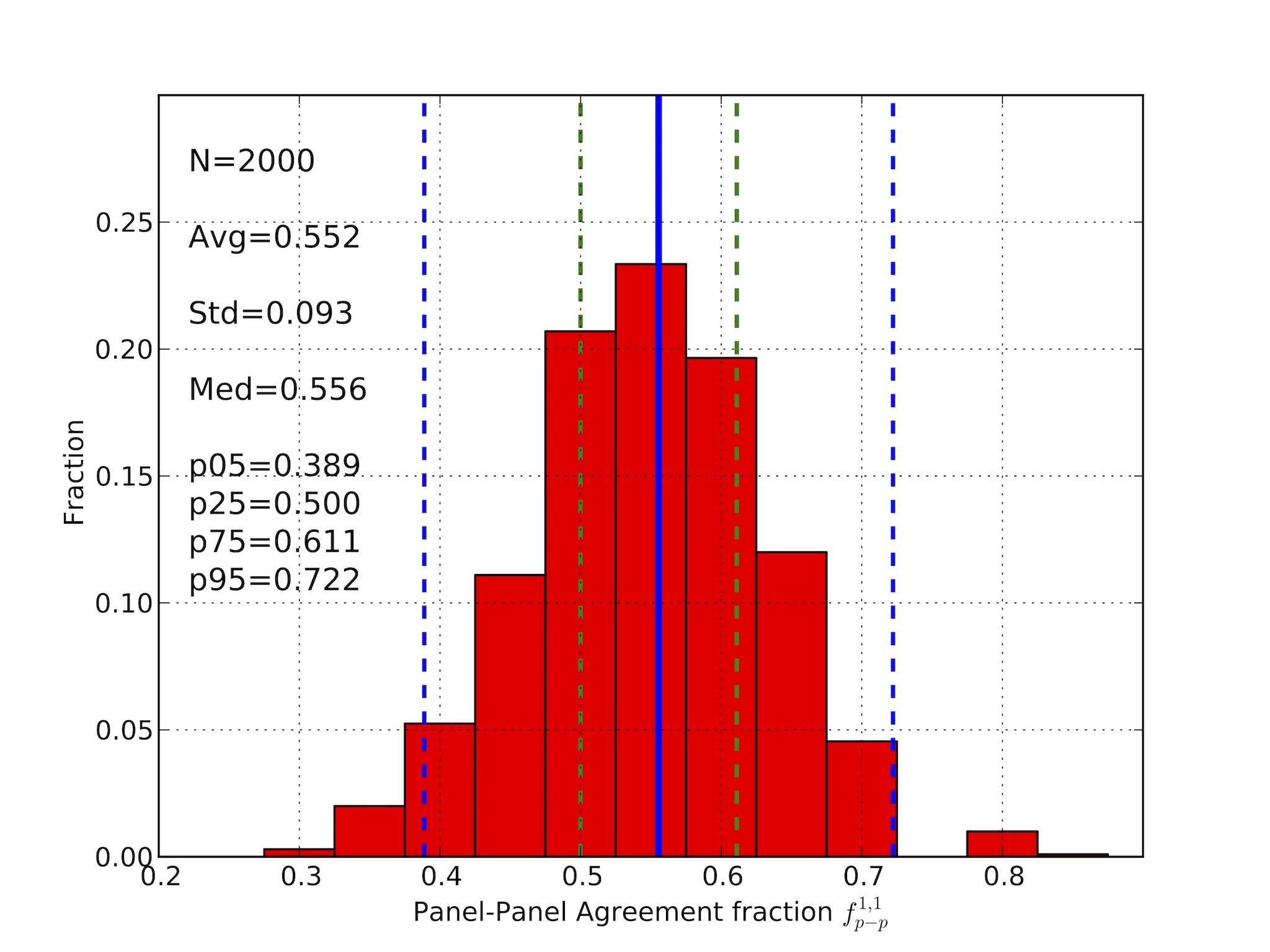}
}
\caption{\label{fig:simpreaf}Distribution of the simulated pre-meeting panel-panel first quartile agreement fraction $f^{1,1}_{p-p}$. The vertical dashed lines indicate the 5th, 25th, 75th and 95th percentiles, while the solid line marks the median value.}
\end{figure}

The simulations show that $f^{k,l}_{p-p}$ displays a roughly Gaussian distribution, with dispersions that range from 4\% to 9\%. An example case for the distribution of $f^{1,1}_{p-p}$ is shown in Figure~\ref{fig:simpreaf}. Finally, the simulated average panel-panel first-quartile correlation $c_{p-p}$ is 0.62$\pm$0.07, to be compared with $c_{r-r}$=0.22$\pm$0.18 for single referee pairs, which provides a quantitative estimate of the improvement produced by having six referees reviewing the same application.

In the TGH context (see Section~\ref{sec:simul}), the reliability of the review process can be quantified computing the agreement fractions with a fictitious panel, the True Panel (TP), which would rank the applications according to their true grades in what we will indicate as the true quartiles (TQ). The agreement matrix for the pre-meeting ranks is presented in Table~\ref{tab:true} for the usual $N_r$=6 case. A comparison to Table~\ref{tab:simppaf} shows that, as expected, the agreement with a TP is better than that expected between two real panels. In other words, single panels are closer to 'truth' than to each other.

These results can be used to answer the following hypothetical question: what are the chances that an application belonging to TQ=$i$ is ranked by a panel in quartile $j$?  One can also consider a milder formulation of the above question: if a proposal is 'objectively' very good, what are the chances that it is ranked in the first quartile by a panel?

According to Table~\ref{tab:true}, a TQ=1 application is ranked in the first quartile about 64\% of the times. This means that, on average, a very good proposal needs to be submitted at least twice to ensure it gets to the top quartile at least once. In the opposite case of a very poor proposal (TQ=4), these chances are very slim, i.e. of the order of a percent (in about 90\% of the cases the proposal ends-up in the two bottom quartiles). The outcome becomes very close to random for TQ=2, for which the chances are just above 25\%. For TQ=3 one needs to re-submit the same proposal about 10 times for being sure to get to the top quartile at least once.

As stated above, these probability estimates are only valid in the TGH context, and assume that the same, unchanged proposal is evaluated by a completely different panel, so that the subjectivity can be described by our model.

\subsection{Outcome expectations}

The model can also give indications on the expectations a principal investigator can have for her/his proposal, when this is submitted to two independent panels (for instance when the proposal is re-submitted after a rejection). One can make predictions for any input TG, similarly to what we did above. However, for a general approach, we discuss the case of an 'average proposal'. For this purpose, we first define the average proposal as characterised by a true grade $\gamma=\gamma_0$ (see Section~\ref{sec:model}). We then simulate the review by the a large number of panel pairs and, for each two-panel session, we derive the two corresponding quartiles and construct what we indicate as the Expectancy Matrix $E$. The generic $E_{i,j}$ element represents the probability that the same proposal is ranked in the $i$-th quartile by panel \#1 ($q_1$) and in the $j$-th quartile by panel \#2 ($q_2)$. 

\begin{table}
\caption{\label{tab:true}Simulated true quartile agreement matrix}
\tabcolsep 4.5mm
\begin{tabular}{c|cccc}
\hline
\hline
 True     & \multicolumn{4}{c}{Panel Quartile}\\
Quartile  & 1 & 2 & 3 & 4\\
\hline
1 & 0.64 &  0.26 & 0.09 &  0.01\\
2 & 0.26 &  0.38 & 0.27 &  0.09\\
3 & 0.09 &  0.27 & 0.38 &  0.26\\
4 & 0.01 &  0.09 & 0.26 &  0.63\\
\hline
\end{tabular}
\end{table}

The results for $N_r$=6 are presented in Table~\ref{tab:expecto}. This reveals that the probability that the two evaluations result into the same quartile ranking is $P(q_1=q_2)\equiv \sum_{i=1}^4 E_{i,i}$=30\%. For instance, if the same identical proposal is resubmitted (unchanged) after a first rejection (and assigned to a different panel) one should expect a consistent outcome in about one third of the cases, leaving room for significant disagreement between the two reviews: $P(q_1>q_2)=P(q_1<q_2)$=35\%.

As expected for an 'average' case, the probability $P(q_1=\{2 \lor 3\} \land q_2=\{2 \lor 3\})$ that the same proposal is ranked in the two central quartiles by both panels is larger: 52\%. 

Finally, the probability that the proposal is ranked in a given quartile by a panel is obtained summing the rows (or the columns) of the Expectancy Matrix. This shows that the average proposal has a non-negligible 14\% chance to end-up in the top quartile (and a similar probability to end-up in the bottom quartile). For $N_r$=3 this increases to 19\%.

\subsection{Panel size}

Since the pre-meeting phase is modelled treating the different referees separately, it is possible to study the effects of varying their number $N_r$. The results of the simulations are shown in Table~\ref{tab:nref} for the first quartile. The table includes also the corresponding Cohen's $k$ values. This allows a further validation test against the direct measurements reported in Section~\ref{sec:subpanel}: the simulated first-quartile agreement fraction for $N_r$=3 is 45\%, which matches rather well the boot-strapped value (43\%).

As expected, the panel-panel agreement increases with the number of referees: it doubles going from one single referee (0.33) to 20 referees (0.70), while changing from 5 to 10 only increases it by a factor 1.17.

\subsection{Number of panels}
\label{sec:npan}

So far we only considered the agreement fractions between two reviewing bodies, be they two distinct referees or two distinct panels grading the same set of proposals. With the model one can go one step further, and ask what is the agreement fraction when one considers a generic set of $N_p$ panels.

The simulations show that, for the adopted value of $\alpha$ and for $N_r$=6, the agreement fraction scales proportionally to $N_p^{-\beta}$, with $\beta\simeq$1. For instance, this is 0.37, 0.29, 0.23 and 0.19 for $N_p$=3, 4, 5 and 6, respectively, while $f_{p-p}<0.1$ for $N_p>$10, and $f_{p-p}<0.01$ for $N_p>$100. Of course these values depend on $N_r$: for instance, $\beta \simeq 0.7$ for $N_r$=10, while $\beta \simeq 1.5$ for $N_r$=3. It is worth noticing that in the case of complete uncorrelation ($\alpha\gg1$), $f_{p-p}=(1/4)^{N_p-1}$, irrespective of $N_r$, which allows a direct comparison with the 'real', partially correlated case.
The simulated agreement fraction  for two panels (with $N_r$=6 and $\alpha$=1.7) is a factor 2.2 larger than the 'random' value. For three panels this ratio is $\sim$6.

\begin{table}
\caption{\label{tab:expecto}Expectancy Matrix for the average proposal}
\tabcolsep 4.5mm
\begin{tabular}{c|cccc}
\hline
\hline
 Panel \#1    & \multicolumn{4}{c}{Panel \#2 Quartile}\\
Quartile  & 1 & 2 & 3 & 4\\
\hline
1 & 0.02 &  0.05 & 0.05 &  0.02\\
2 & 0.05 &  0.13 & 0.13 &  0.02\\
3 & 0.05 &  0.13 & 0.13 &  0.05\\
4 & 0.02 &  0.05 & 0.05 &  0.02\\
\hline
\end{tabular}
\end{table}

\section{Discussion}
\label{sec:disc}

Although a number of weaknesses were identified (see \cite{roy85} and \citet{smith} for critic summaries), peer-review encounters the support of the majority of scientists, and is hence widely used as a deciding mechansim for allocating resources. In general, peer-review can be seen as a democratic process, as theorised by \cite{kitcher}.
In this approach, which Kitcher calls 'well ordered science', the science that society should choose to carry out and support is the science that is most favoured by the scientific community. The panelists serve as representatives of that community, and they are being asked to use the knowledge and skills they have acquired over their careers to assess what they judge to be the most productive applications. A different panel would produce a different selection, which would have a similar merit and carry the promise of producing an equivalent scientific output.

In the context of the TGH introduced in this paper (see Section~\ref{sec:model}), this is equivalent to the result that two different panels produce, on average, two selections that are equally close to that provided by an ideal True Panel.

That said, the limited reproducibility and the possible introduction of selection biases are often used against peer-review as a concept. For instance, \cite{gillies} argues that a random selection would be better, because it avoids what he calls the 'systemic bias', which tends to 'favor mainstream research programmes', leading to 'the stifling of new ideas and of innovation'. Other problems of peer-review are related to the progressively degrading quality of the reviews in connection to the increasing review load (see for instance \cite{fox}; \cite{bohannon}). Additional reported limitations are low predictive power, potential conservatism and risk aversion (see the general introduction in \cite{snell} and references therein).

\begin{table}
\caption{\label{tab:nref}Panel-panel first quartile agreement fraction}
\tabcolsep 3.6mm
\begin{tabular}{ccccccc}
\hline
\hline
$N_{r}$ & $f_{p-p}$ & $k$ & & $N_{r}$ & $f_{p-p}$ & $k$\\
\hline
1 & 0.33 &0.11 & & 7 & 0.57 & 0.43\\
2 & 0.39 &0.19  && 8 & 0.59& 0.45\\
3 & 0.45 &0.27  && 9 & 0.60& 0.47\\
4 & 0.48 &0.31  &&10 & 0.62& 0.49\\
5 & 0.53 &0.37  &&20 & 0.70 & 0.60\\
6 & 0.54 &0.39  &&100 & 0.86& 0.81\\
\hline
\end{tabular}
\end{table}

In this paper we focused on the quantitative characterisation of the subjective part of the pre-meeting process, which plays an important role in the way the community perceives and judges the reliability of the process. Although applicants base their evaluation of the peer-review paradigm on their own experience, and normally do not have means of estimating how much the selection is really subjective, this perception is increasingly diffused and, in our experience, creates a significant level of frustration. This is normally coupled to the level of feedback that unsuccessful applicants get, judged as unsatisfactory and, at times, as provided by 'incompetent' referees. For these reasons, it is important to estimate the level of reproducibility of the selection process, in order to raise the awareness in the community about its inherent limitations.
On the other hand, the quantification of the subjective part is fundamental for the functioning of the whole process. In their very recent study, \cite{pluchino} warn against the risk of what they call 'naive meritocracy', in which the underestimate of the role of randomness leads to a failure in properly rewarding intrinsically meritorious cases.

For obvious reasons there are not very many quantitative studies that try to assess the repeatability of the selections operated by a given panel. \cite{cole} reported the results of an experiment in which 150 proposals submitted to the National Science Foundation were evaluated independently by a second set of reviewers, and concluded that 'the fate of a particular grant application is roughly half determined by the characteristics of the proposal and the principal investigator, and about half by apparently random elements which might be characterised as the "luck of the reviewer draw"'. 

A more recent study is the NIPS experiment \citep{cortes}. The organisers of the Neural Information Processing Systems (NIPS) conferences, an important series in theoretical computer science, assigned 10\% of the applications (166) to two different panels, without the panel members knowing which ones were in common. The two committees were tasked with a 22.5\% acceptance rate. The final result was that 57\% of the papers accepted by the first panel were rejected by the second (and vice versa), with an estimated 95\% confidence level of 40-75\% \citep{price}. In our terminology and within the same confidence level, this corresponds to a panel-panel agreement fraction (within percentile 22.5) of 43$\pm$18\%. 

The NIPS papers were typically reviewed by three referees. Assuming that these behave in the same way as those discussed in this study, we can use our model to reproduce that experiment. The simulations yield a median agreement fraction within percentile 22.5 of 43.2$\pm$13.5\% (at the 95\% confidence level), with a standard deviation of 6.8\%. This is fully consistent with the results derived from the NIPS experiment, and provides a direct confirmation of the reliability of our model. Incidentally, this is also well in line with our direct bootstrapping results for the first quartile reported in Section~\ref{sec:subpanel} for sub-panels composed by 3 reviewers (43\%), which confirms that the statistical behaviour of the referees in our sample is comparable to that of the completely different context of the NIPS sample.

The data presented here show that the intersection between the lists of programmes ranked in the first quartile by two independent referees is 34$\pm$28\% (95\% confidence level), while this is consistent (within the noise) with a purely random process in the central quartiles (25\%). The situation improves when considering a panel: for 6 members, the first-quartile agreement is on average 55$\pm$17\%, while this reduces to $\sim$33$\pm$16\% in the second and third quartiles, which is only marginally above the random limit. These findings are along the lines presented by \cite{snell}, who reports increased random effects in the mid-range of rankings. For this reason, he suggests that top- and bottom-ranked proposals do not need further discussion, which should only take place for the mid-range cases, be it asynchronous online or face-to-face. 

In the context of the TGH (see Section~\ref{sec:model}), increasing the number of referees does improve the reproducibility of the process and, therefore, one would want to enlarge the size of the panels. However, especially when these have to be convened in physical meetings, this is hard to manage, both in terms of logistics and costs. There is no obvious criterion as to what the optimal number of reviewers is. One could tentatively set it by imposing a minimum panel-panel agreement fraction. For instance, setting this limit to 50\%, from our simulations one would conclude that panels should include 4-5 referees (see Table~\ref{tab:nref}). Based on similar considerations, \cite{snell} reached an analogous conclusion: "five reviewers per application represent a practical optimum which avoids large random effects evident when fewer reviewers are used, a point where additional reviewers at increasing cost provides only diminishing incremental gains in chance-corrected consistency of decision outcomes."

\vspace{2mm}
We will come back to these topics in Paper II, in which we will present and analyse the effects of the panel discussions, provide quantitative estimates on the final agreement fractions, and discuss the usefulness of the meetings.

\vspace{5mm}

\section{Conclusions}
\label{sec:conc}

In this paper we presented a statistical analysis of the pre-meeting proposal evaluation data for $\sim$15000 telescope time applications for ESO telescopes, reviewed by $\sim$500 referees, who assigned over 140000 grades in about 200 panel sessions. The main results can be summarised as follows:

\begin{enumerate}
\item The grade distributions of the single referees show a great variety and display marked deviations from a Normal behaviour. 
\item Several correlations between the parameters characterising the distributions are detected. In particular, the width of the distributions tends to be larger for larger central values.
\item On average, only about one third of the proposals ranked in the top quartile by one referee are ranked in the same quartile by any other referee of the panel. A similar value is observed for the bottom quartile.
\item In the two central quartiles, the agreement fractions are very marginally above the value expected for a fully aleatory process (25\%).
\item The average panel-panel agreement fraction measured bootstrapping the available data for sub-panels composed by 3 referees is 43\% in the first quartile and 46\% in the fourth quartile, while in the two central quartiles this is 30\%, i.e. marginally above the random limit (Cohen's $k$=0.07).
\item A model based on the observed statistical properties of single referees and on the assumption that a True Grade can be assigned to a proposal gives a satisfactory description of the pre-meeting phase.
\item The model, applied to the ESO case ($N_r$=6 reviewers per panel), predicts that the expected first and fourth quartile panel agreement fraction is 55$\pm$17\% (95\% confidence level). The corresponding fraction for the central quartiles is 33$\pm$16\%.
\item The data and the simulations confirm the diffused perception that top and bottom cases are easier to identify, whilst the process becomes fuzzier in the central quartiles.
\item When $N_r$ is reduced to 3, the average first/fourth quartile agreement fraction is 45$\pm$19\%. The agreement fraction exceeds 50\% for $N_r\geq$5.
\item The model reproduces both the results deduced from bootstrapping the data in sub-panels including 3 referees, and the NIPS experiment, in which the same set of applications was assigned to two different panels.
\item The pre-meeting phase can be characterised by one single parameter $\alpha$, defined as the ratio between the dispersion of the grading process and the intrinsic dispersion of the True Grade. The best match with the data is achieved for $\alpha$=1.7.
\item The typical SIQR rank uncertainty for a panel with $N_r$=6 is about 32\% in the mid-rank region, and about 8\% at either edge of the rank range.
\item Referee calibration does not change significantly the quartile classification operated by a panel. The subjectivity inherent to the review process dominates over the systematic differences characterising the reviewers. 
\end{enumerate}

\begin{acknowledgements}
This paper is fruit of independent research and is not to be considered as expressing the position of the European Southern Observatory on proposal review and telescope time allocation procedures and policies.

The author is indebted to Elisabeth Hoppe, for the kind and enduring assistance provided throughout this research project, to Gaitee Hussain, Francesca Primas and Dimitri Gadotti for fruitful discussions on proposal review matters, and to Eric Price for providing the information on the NIPS experiment and his expert advice. The author is particularly grateful to the referee, Neill Reid, for pointing him to the work by Philip Kitcher and for the very useful suggestions, which greatly helped improving the quality and the clarity of this paper.
\end{acknowledgements}




\appendix

\section{Referee calibration revisited}
\label{sec:calib}

In this section we will approach the problem of referee calibration (see Section~\ref{sec:norm}), in the light of the data and the model presented in this paper. Since different panel sessions have different numbers of proposals assigned, for convenience we introduce the normalized rank:

\begin{displaymath}
\bar{r} = \frac{r-1}{N-1}
\end{displaymath}

where $N$ is the number of runs reviewed in the given panel session and $r$=1, 2, ..., $N$ is the run rank. The normalized rank is 0 for the top- and 1 for the bottom-ranked run, respectively.

\subsection{Effects of referee calibration}

\begin{figure} 
\centerline{
\includegraphics[width=10cm]{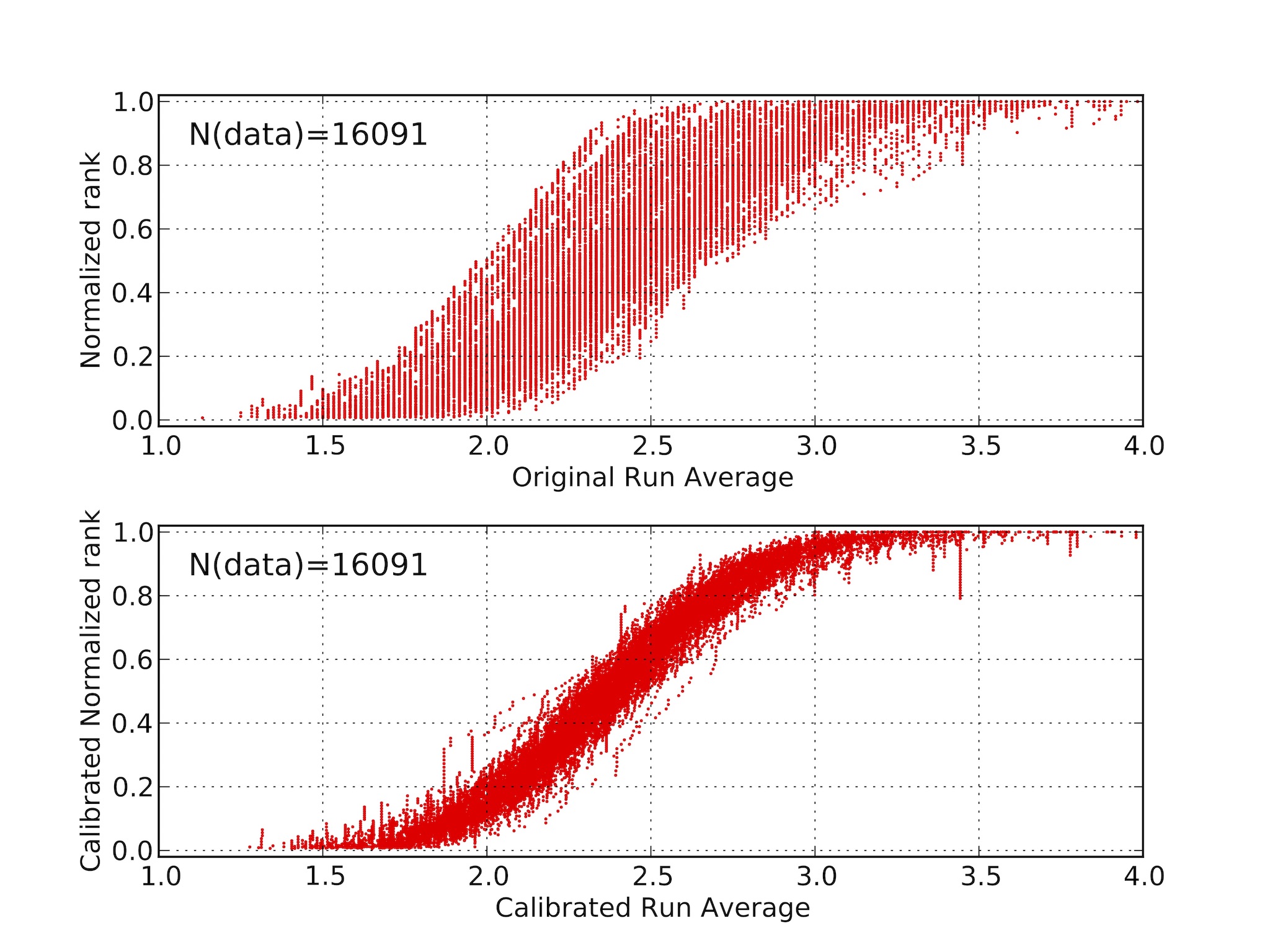}
}
\caption{\label{fig:precalib}Correlation between run average grade and normalized ranking for uncalibrated (upper panel) and calibrated (lower panel) pre-meeting data with $N_r$=6.}
\end{figure}

In Figure~\ref{fig:precalib} we present the correlations between average run grade and normalized rank for calibrated and uncalibrated pre-meeting data. For the calibration we have used the prescription described in Section~\ref{sec:norm}, so that after applying it the grade distributions of the 1223 referee sessions all have the same average and standard deviation. Only runs with $N_r$=6 referees were included.

The calibration clearly improves the correlation between the average grade and the corresponding rank. This demonstrates that, although residual scattering is present, the calibration is effective in terms of homogenising the scales of the different referees. It is therefore important when grades given by different panels to different runs are merged together to form a single ranked list (as is the case when different panels review different sets of proposals to be allocated at the same telescope).

Since, in the end, what matters for the fate of a run is its ranking, it is important to examine the effect that calibration has on it. This is illustrated in Figure~\ref{fig:normrank}, where we plotted the calibrated normalized rank (in the given panel) as a function of the original (uncalibrated) normalized rank. The correlation is very strong, especially in the first and in the last quartiles, implying that the calibration does not produce a dramatic change in the rank of a run.

This can be quantified through what we will indicate as quartile change matrix, which is presented in Table~\ref{tab:ranknorm} for the same data set used above. This clearly shows that the agreement between original and calibrated ranks is far better than what is measured between single referees (see Table~\ref{tab:ira}) or between sub-panels reviewing the same set of runs (see Sec.~\ref{sec:subpanel}).

\subsection{Does referee calibration really help?}

Obviously, the calibration does not change the rank of a run within the set assigned to a given referee. Precisely because of this reason, calibration does not have any effect on the referee-referee agreement fraction. However, as we have shown in the previous section, the rank constructed from the run average does change. The question is whether calibration increases the agreement between two different panels reviewing the same set of proposals. In the context of the TGH the question can be reformulated as follows: does referee calibration improve the match with the true rank?

This question can be addressed directly with the available data, applying the bootstrap approach described in Sec.~\ref{sec:subpanel} to sub-panels formed by 3 referees. This is simply done comparing the panel-panel quartile agreement matrix derived using the original or the calibrated referee grades. The calculations show that these are statistically indistinguishable.

\begin{table}
\caption{\label{tab:ranknorm}Quartile change matrix for original and calibrated runs}
\tabcolsep 4.5mm
\centerline{ 
\begin{tabular}{c|cccc}
\hline
\hline
   original & \multicolumn{4}{c}{calibrated quartile}\\
quartile  & 1 & 2 & 3 & 4\\
\hline
1 & 0.79 &  0.19 & 0.02 &  0.00\\
2 & 0.21 &  0.52 & 0.24 &  0.03\\
3 & 0.01 &  0.28 & 0.51 &  0.20\\
4 & 0.00 &  0.01 & 0.22 &  0.77\\
\hline
\end{tabular}
}
\end{table}

As for the second version of the question, this can be addressed using our model, and deriving the agreement fraction between the simulated quartile and the TQ (see Sec.~\ref{sec:simpre}). The simulations show that there is no statistically significant improvement in the agreement fractions when the referee calibration is applied.

Therefore, the reason for the lack of improvement when applying referee calibration resides in the fact that the stochastic component affecting the single ranks dominates over the systematic effect related to the different grading scales. This confirms our previous conclusion that calibration is only relevant when merging lists of different proposals graded by different panels, to homogenise their grade scales. Obviously, the need for calibration goes away if the panels are tasked with using an ordinal (rank) rather than a cardinal (grade) procedure to rate the applications.

From this we conclude that, although the calibration changes the rank of a given run within the set assigned to a given panel, the process does not increase the agreement with an hypothetical, different panel reviewing the same set of applications, nor does it improve the match with the TQ.

Further insights into the reasons why this is so can be gained examining one last set of simulations in which, for a large number of runs, we have derived the pre-meeting rank as a function of the input true rank, with the aim of deriving the associated uncertainty. 
The calculations show that, while the median rank provides a close proxy to the true rank, it is accompanied by a large dispersion. More precisely, this dispersion is definitely larger than that observed in the relation between original and calibrated ranks.

\begin{figure} 
\centerline{
\includegraphics[width=10cm]{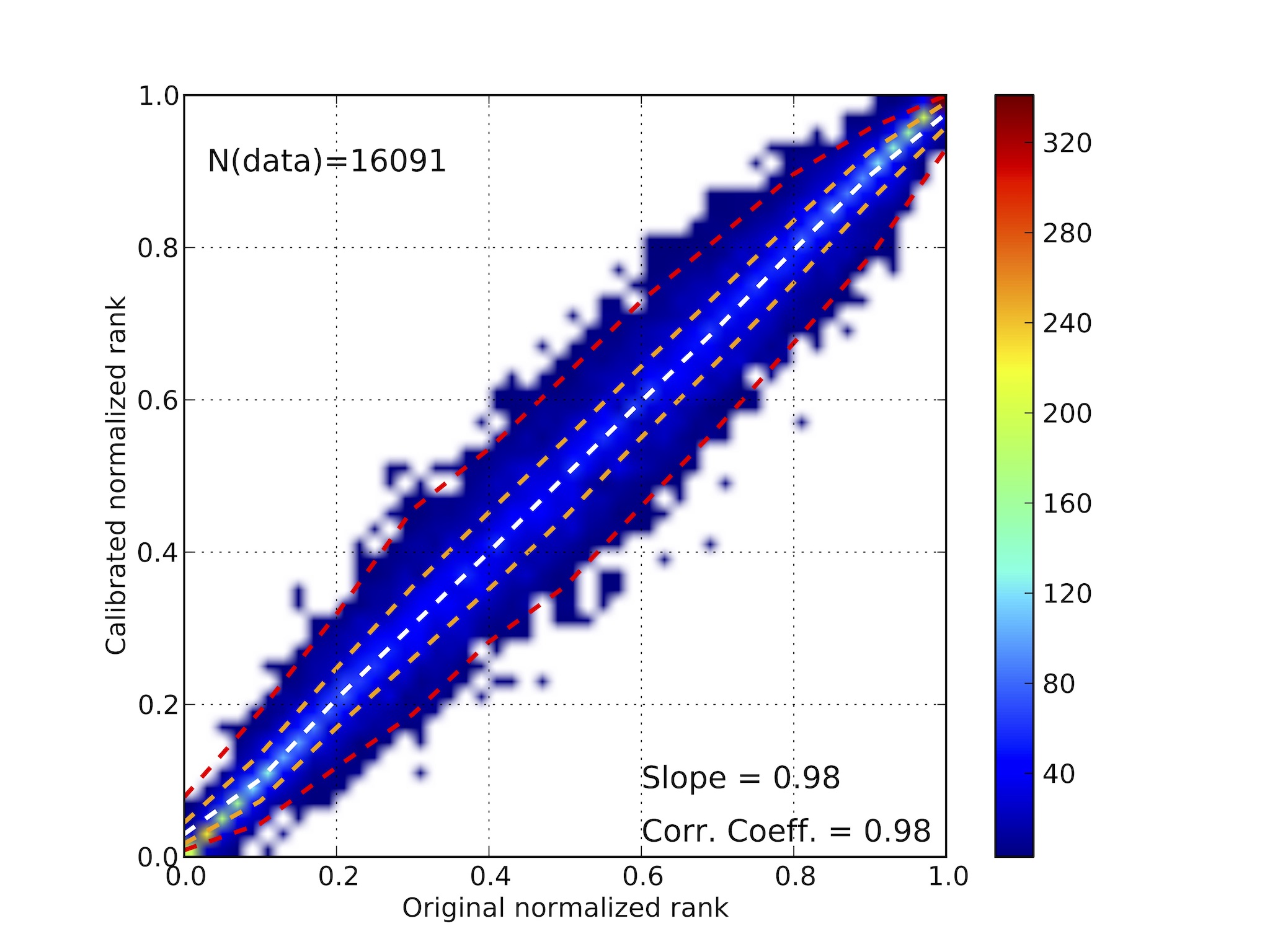}
}
\caption{\label{fig:normrank}Correlation between original and calibrated rank for the pre-meeting data with $N_r$=6. The dashed lines indicate the 5-th (red), 25-th (orange), 50-th (white), 75-th (orange), and 95-th (red) percentiles (bottom to top).}
\end{figure}

\begin{figure} 
\centerline{
\includegraphics[width=10cm]{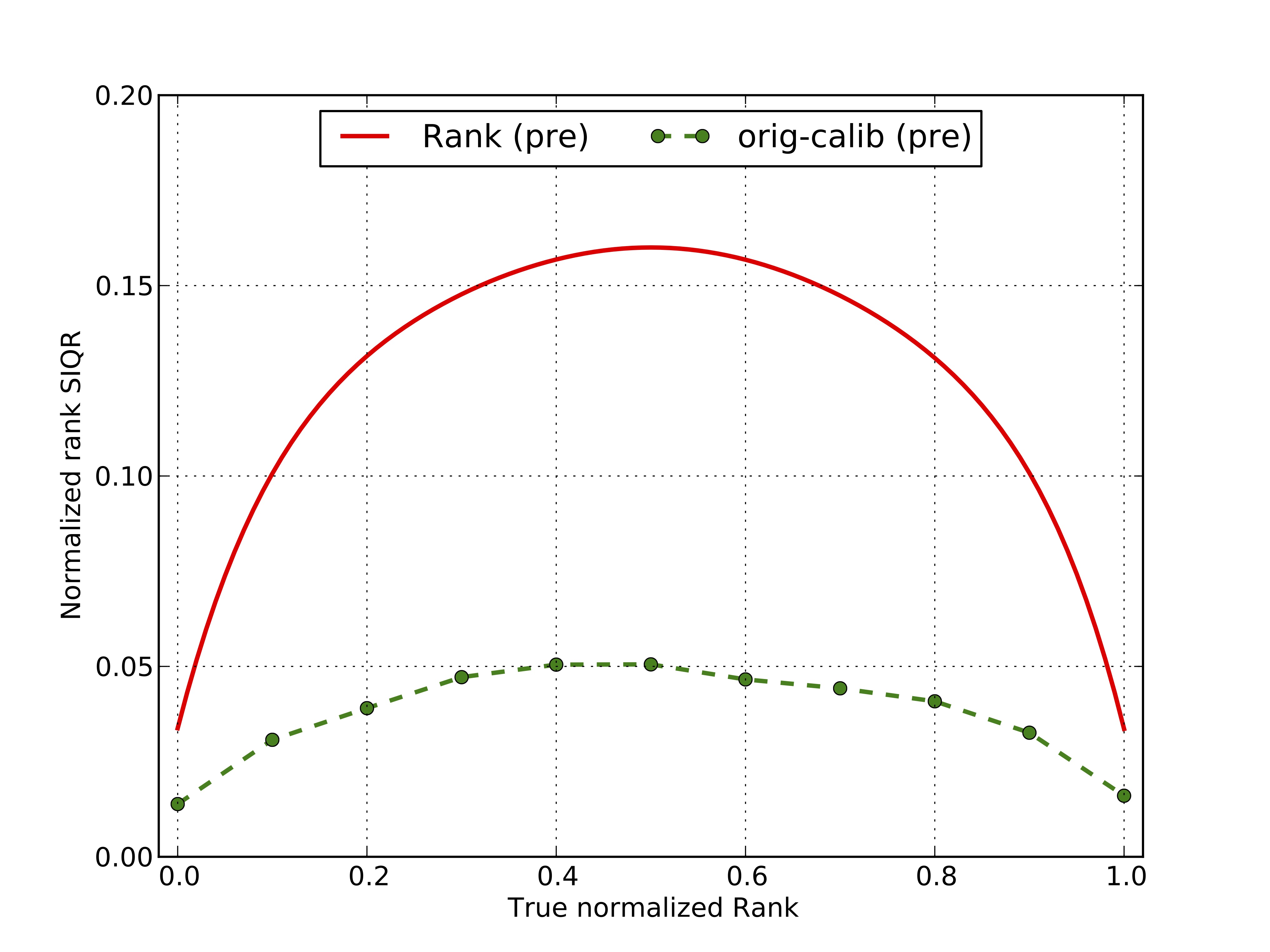}
}
\caption{\label{fig:ranksiqr} Simulated semi-interquartile range uncertainty of the pre-meeting normalized rank as a function of the input true rank (red), for $N_r$=6. For comparison, the SIQR rank change produced by referee calibration in the pre-meeting data (green) is also plotted.}
\end{figure}

The expected uncertainty on the pre-meeting normalized rank can be quantified with the SIQR, which is plotted in Figure~\ref{fig:ranksiqr}. The normalized SIQR is about 0.04 at the edges of the rank range, and it steadily grows towards the centre, where it reaches about 0.16. For example, in a panel that was assigned 70 runs, the rank uncertainty (at the 50\% confidence level) for mid-ranked runs is $\pm$11. This increases to $\pm$24 at the 90\% confidence level. For comparison, in Figure~\ref{fig:ranksiqr} we have plotted the measured SIQR rank change produced by the pre-meeting data calibration (see also Figure~\ref{fig:normrank}). This is typically 3 times smaller than the rank uncertainty. For the same panel assignment, the rank variation introduced by referee calibration in the mid-rank region is only about $\pm$3 (SIQR).

\section{The agreement fractions}
\label{sec:appa}

Let $R=\{r_1, r_2, ..., r_{n(R)}\}$ be the set of $n(R)$ runs assigned to the given panel (in this appendix we will indicate with $n(X)$ the cardinality of set $X$), and $f_i(g)$ be the grade distribution of the $i$-th referee, with $i=1, 2, ..., N_r$, where $N_r$ is the number of referees that compose the panel. We then introduce the two threshold grades $g_{p_1,i}$ and  $g_{p_2,i}$  for the $i$-th referee, defined as:

\begin{displaymath}
\int_1^{g_{p_1,i}} f_i(x) \; dx =p_1 \; \int_1^5 f_i(x) \; dx
\end{displaymath}

and 

\begin{displaymath}
\int_1^{g_{p_2,i}} f_i(x) \; dx =p_2 \; \int_1^5 f_i(x) \; dx
\end{displaymath}

where $p_1$ and $p_2$ ($0\leq p_{1/2} \leq 1$) are two input parameters that set the portion of the distribution one wants to use for computing the agreement. For instance, if $p_1$=0.0 and $p_2$=0.25, the runs with $g_{p_1,i}\leq g_i< g_{p_2,i}$ are those which were ranked by the $i$-th referee in the first quartile of her/his distribution. Similarly, if $p_1$=0.75 and $p_2$=1.0, the selected runs are those ranked in the fourth quartile. 

Let then $g_i(r)$ be the grade attributed by the $i$-th referee to run $r$, and $p=(p_1,p_2)$ the selection range. With these positions we can introduce the run selection set for the $i$-th referee:

\begin{equation}
\label{eq:set}
R_{p,i}  =   \{  r \in R   \mid  g_{p_1,i} \leq g_i(r) < g_{p_2,i} \}
\end{equation}

\subsection{Referee-referee agreement fraction}
\label{sec:ira}

With the above positions, one can easily introduce the referee-referee (r-r) agreement, which we define as follows:

\begin{equation}
f_{i,j} = \frac{n( R_{p,i} \cap R_{p,j})}{n(R_{p,i})} \;\;\; \text{with} \;\;\; 0\leq i < N_r \text{,}\; \; i < j \leq N_r
\end{equation}

which is the fraction of runs selected by the $i$-the referee that are in common with the selection operated by the $j$-th referee of the given panel. For each panel session the above definition produces $N_r (N_r-1)/2$ agreement values. The calculation can be extended to all panel sessions, and the r-r agreement distribution can be readily derived.

\subsection{Referee-majority agreement fraction}
\label{sec:rm}

The number of referees in the panel that selected (according to Eq.~\ref{eq:set}) the given run $r_j \in R_{p,i}$ is given by

\begin{displaymath}
n_{i,j} = \sum_{k=1}^{N_r} n\left( \{ r_j\} \cap R_{p,k} \right)
\end{displaymath}

where it is $1\leq n_{i,j}\leq N_r$. With this position, the set of runs selected by referee $i$ that were also selected by at least $m$ referees is:

\begin{displaymath}
S_{m,i} = \{r_j \in R_{p,i} \mid n_{i,j} \geq m \}
\end{displaymath}

This finally leads to the definition of the agreement fraction for the $i$-th referee:

\begin{equation}
f_{m,i} = \frac{n(S_{m,i})}{n(R_{p,i})}
\end{equation}

This figure of merit can be computed for all referees of all panels, to derive the average agreement fraction $f_m$. This provides an estimate of the typical fraction of runs selected by a referee that are also selected by at least $m$ referees.

For the trivial case in which $m$=1, it is obviously always $f_{m,i}$=1. For $m$=$N_r$, $f_m$ is the fraction of runs that were selected, by all panel members (unanimous agreement). In our specific case ($N_r$=6), we decided to use $m$=4, as this is equivalent to counting the occurrences that would be selected by the majority of the panel (50\%+1 of the members) if they were to vote based on the grades they attributed to the runs. For this reason we will call it referee-majority (r-m) agreement fraction.

\section{Truncated Normal distributions}
\label{sec:truncated}

Although in our simulations we have used the observed distributions, we have run an extensive set of tests modelling the referees with truncated Gaussians, which provide a simple parametric description of real data.

A truncated normal distribution is produced by an underlying normal distribution with average $\mu$ and standard deviation $\sigma$, and bounded in the interval $(a,b)$. It is important to make a distinction between the underlying $\mu$ and $\sigma$, and the corresponding parameters of the truncated distribution, which we will indicate as $\mu_m$ and $\sigma_m$. Let us introduce the following variables:

\begin{displaymath}
\alpha= \frac{a-\mu}{\sigma} \;\; \mbox{and} \;\; \beta=\frac{b-\mu}{\sigma}
\end{displaymath}

If $\phi(x)$ and $\Phi(x)$ are the normal pdf and its cdf, respectively, after introducing the quantity $Z=\Phi(\beta)-\Phi(\alpha)$, $\mu_m$ and $\sigma_m$ can be computed as follows (see for instance \cite{trunc}):

\begin{equation}
\label{eq:mu}
\mu_m = \mu + \frac{\phi(\alpha)-\phi(\beta)}{Z}
\end{equation}
\begin{equation}
\label{eq:sig}
\sigma_m^2 = \sigma^2 \left [  1 + \frac{\alpha\; \phi(\alpha) - \beta\; \phi(\beta)}{Z}  - 
\left(  \frac{\phi(\alpha)-\phi(\beta)}{Z} \right)^2 \right ]
\end{equation}

\begin{figure} 
\centerline{
\includegraphics[width=10cm]{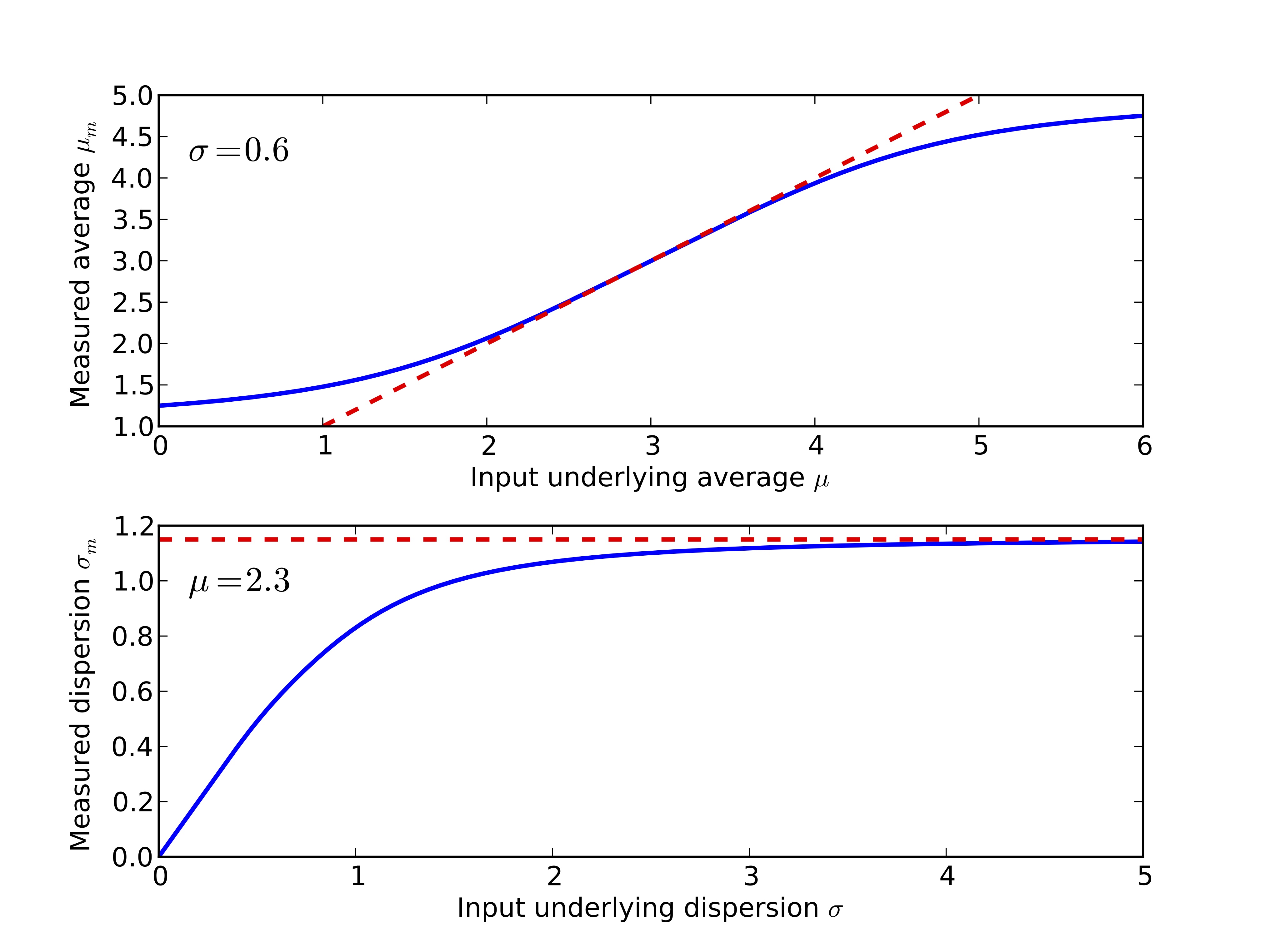}
}
\caption{\label{fig:trans}Example transformations between underlying and measured parameters for a truncated normal distribution with $a=1$ and $b=5$. Upper panel: measured average as a function of underlying average $\mu$ for a constant $\sigma$=0.6. The dashed line traces the $\mu_m=\mu$ relation.
Lower panel: measured standard deviation as a function of underlying $\sigma$ for a constant $\mu$=2.3. The dashed horizontal line indicates the limit value $\sigma_m=1.15$ (see text).}
\end{figure}

These equations allow one to predict what the 'observed' average and variance will be for a given input combination of the two parameters. Example transformations are presented in Figure~\ref{fig:trans}, for the specific case of $a=1$ and $b$=5. The upper panel illustrates the deviation from linearity in the relation between $\mu$ and $\mu_m$ as one approaches the truncation boundaries, for the example case with $\sigma$=0.6: the measured average will never be smaller than $\sim$1.25 or larger than $\sim$4.75 (for this particular choice of $\sigma$). The range of measured averages becomes progressively smaller for increasing values of $\sigma$. For $\sigma>3$ the measured average remains almost constant around the center of the truncation interval (3.0 in our specific case).

The lower panel shows another important aspect, i.e. the saturation of $\sigma_m$ for progressively larger values of the underlying $\sigma$. The relation between the two parameters deviates from linearity for $\sigma>0.6$, and approaches the asymptotic value for $\sigma>3$, which corresponds to an approximately flat distribution within the truncation boundaries. In these circumstances, the measured $\sigma_m$ can be computed as:

\begin{displaymath}
\sigma_m^2= \int_a^b (x-\mu_m)^2 \; \phi(x)\; dx
\end{displaymath}

For the limiting case of a flat distribution ($\phi(x)=1/4$ and $\mu_m$=3.0 for the specific truncation boundaries), this translates to $\sigma_m=\sqrt{\frac{4}{3}}\approx$1.15. This implies that any increase in the underlying  $\sigma$ parameter does not produce any increase in the observed dispersion. Incidentally, this limit value is consistent with what is observed in the real data, where $\sigma<1.2$ (see Figure~\ref{fig:summary}, upper right plot). The practical consequence is that, for instance, in order to generate a simulated distribution with $\sigma_m$=1.0 and $\mu_m$=2.3, one needs to input $\sigma\sim1.6$.

For our test simulations, in which we need to reproduce the observed values of $\mu_m$ and $\sigma_m$, we precomputed a fine grid  of $(\mu,\sigma)-(\mu_m,\sigma_m)$ pairs with Equations~\ref{eq:mu} and \ref{eq:sig}, and used it to numerically invert the transformation. This was achieved by finding the nearest neighbour of the measured parameters in the grid, and by retrieving the corresponding underlying parameters to be entered in the random generator routine.

\begin{figure} 
\centerline{
\includegraphics[width=10cm]{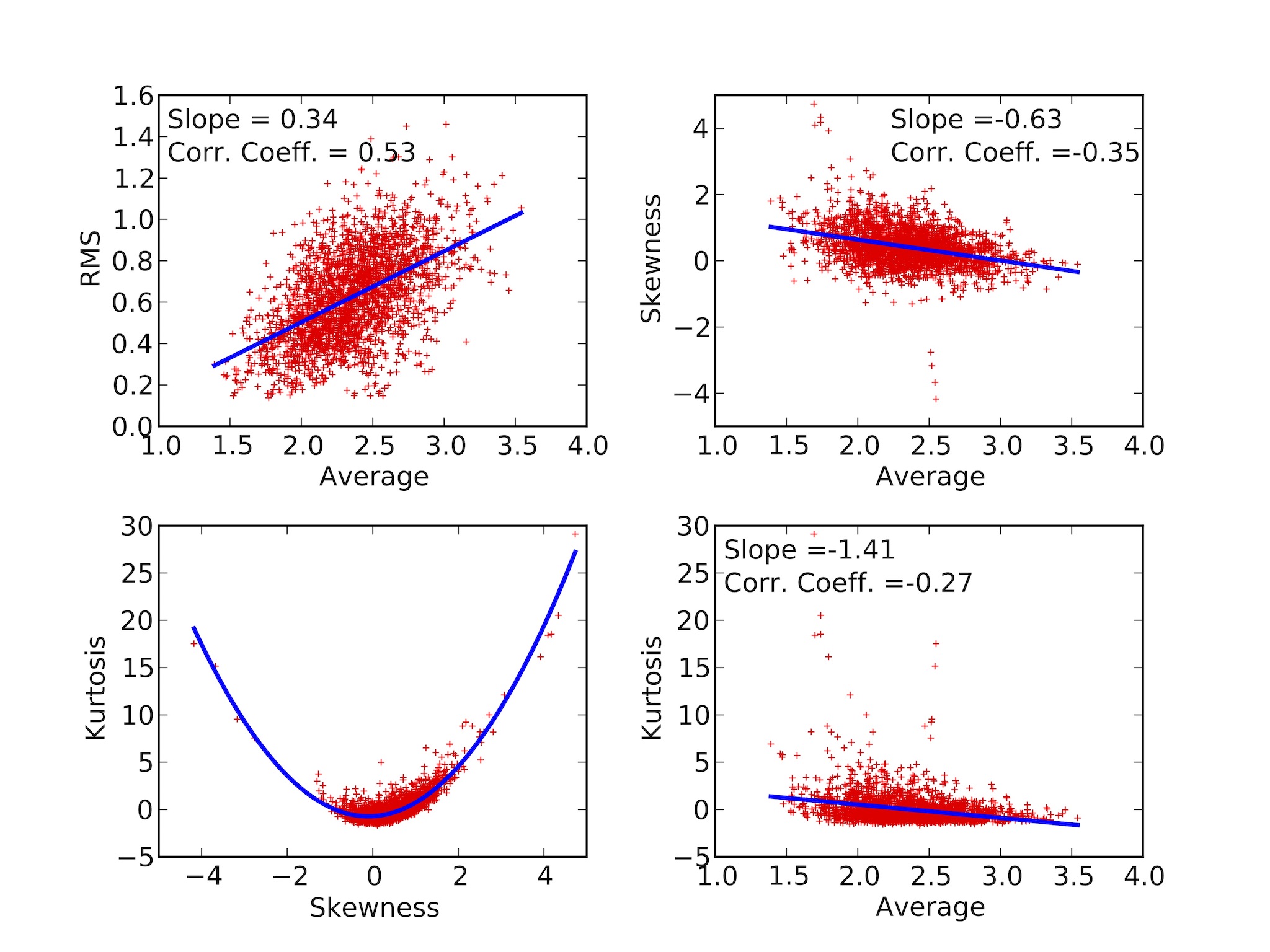}
}
\caption{\label{fig:simcorr} Correlations between the global statistical estimators for pre-meeting simulated grade distributions.}
\end{figure}

In the test simulations, $\mu_m$ is randomly generated using a truncated Gaussian with $\sigma$=0.6, as deduced from the data. Then, $\sigma_m$ is computed using the observed relation $\sigma_m=-0.19 + 0.35 \; \mu_m$ (see Equations~\ref{eq:general}), to which a normal noise with the observed dispersion (0.21; see Figure~\ref{fig:summary})  is added. This completely defines the grade distributions of the referees, and it does not leave any free parameter. The simulations show that this is sufficient to qualitatively reproduce the observed correlations and distributions of the global statistical estimators. For instance, the simulated distributions show the same observed systematic tendency to have a negative Kurtosis (-0.34 vs. -0.27 observed) and a positive Skewness (0.23 vs. 0.40 observed). Similarly, the synthetic distributions show the observed negative correlation between average and Skewness, and the parabolic relation between Skewness and Kurtosis. This is illustrated in Figure~\ref{fig:simcorr}, which can be directly compared to Figs.~\ref{fig:avesig} and \ref{fig:kurskw}.

Considering that there is no free parameter to control the third and fourth moments of the distributions, truncated Gaussians provide a reasonable, fully constrained description of the observed grade distributions. More refined parameterised treatments will require distributions that offer the possibility of tuning their Skewness and Kurtosis.

\end{document}